\documentclass[twocolumn, twocolappendix]{aastex63}

\usepackage{amsmath}
\usepackage{txfonts}
\usepackage{threeparttable}
\hypersetup{allcolors=blue}

\begin{document}
\title{The Cosmic Evolution of the Supermassive Black Hole Population: A Hybrid Observed Accretion and Simulated Mergers Approach}

\author[0000-0002-4436-6923]{Fan Zou}
\affiliation{Department of Astronomy, University of Michigan, 1085 S University, Ann Arbor, MI 48109}
\affiliation{Department of Astronomy and Astrophysics, 525 Davey Lab, The Pennsylvania State University, University Park, PA 16802, USA}
\affiliation{Institute for Gravitation and the Cosmos, The Pennsylvania State University, University Park, PA 16802, USA}

\author[0000-0002-0167-2453]{W. N. Brandt}
\affiliation{Department of Astronomy and Astrophysics, 525 Davey Lab, The Pennsylvania State University, University Park, PA 16802, USA}
\affiliation{Institute for Gravitation and the Cosmos, The Pennsylvania State University, University Park, PA 16802, USA}
\affiliation{Department of Physics, 104 Davey Laboratory, The Pennsylvania State University, University Park, PA 16802, USA}

\author[0000-0001-5802-6041]{Elena Gallo}
\affiliation{Department of Astronomy, University of Michigan, 1085 S University, Ann Arbor, MI 48109}

\author[0000-0002-9036-0063]{Bin Luo}
\affiliation{School of Astronomy and Space Science, Nanjing University, Nanjing, Jiangsu 210093, People's Republic of China}
\affiliation{Key Laboratory of Modern Astronomy and Astrophysics (Nanjing University), Ministry of Education, Nanjing 210093, People's Republic of China}

\author[0000-0002-8577-2717]{Qingling Ni}
\affiliation{Max-Planck-Institut f\"{u}r extraterrestrische Physik (MPE), Gie{\ss}enbachstra{\ss}e 1, D-85748 Garching bei M\"unchen, Germany}

\author[0000-0002-1935-8104]{Yongquan Xue}
\affiliation{CAS Key Laboratory for Research in Galaxies and Cosmology, Department of Astronomy, University of Science and Technology of China, Hefei 230026, China}
\affiliation{School of Astronomy and Space Sciences, University of Science and Technology of China, Hefei 230026, China}

\author[0000-0002-6990-9058]{Zhibo Yu}
\affiliation{Department of Astronomy and Astrophysics, 525 Davey Lab, The Pennsylvania State University, University Park, PA 16802, USA}
\affiliation{Institute for Gravitation and the Cosmos, The Pennsylvania State University, University Park, PA 16802, USA}

\email{E-mail: fanzou01@gmail.com}

\begin{abstract}
Supermassive black holes (SMBHs) can grow through both accretion and mergers. It is still unclear how SMBHs evolve under these two channels from high redshifts to the SMBH population we observe in the local universe. Observations can directly constrain the accretion channel but cannot effectively constrain mergers yet, while cosmological simulations provide galaxy merger information but can hardly return accretion properties consistent with observations. In this work, we combine the observed accretion channel and the simulated merger channel, taking advantage of both observations and cosmological simulations, to depict a realistic evolution pattern of the SMBH population. With this methodology, we can derive the scaling relation between the black-hole mass ($M_\mathrm{BH}$) and host-galaxy stellar mass ($M_\star$) and the local black-hole mass function (BHMF). Our scaling relation is lower than those based on dynamically measured $M_\mathrm{BH}$, supporting the claim that dynamically measured SMBH samples may be biased. We show that the scaling relation has little redshift evolution. The BHMF steadily increases from $z=4$ to $z=1$ and remains largely unchanged from $z=1$ to $z=0$. The overall SMBH growth is generally dominated by the accretion channel, with possible exceptions at high mass ($M_\mathrm{BH}\gtrsim10^{8}~M_\odot$ or $M_\star\gtrsim10^{11}~M_\odot$) and low redshift ($z\lesssim1$). We also predict that around 25\% of the total SMBH mass budget in the local universe may be locked within long-lived, wandering SMBHs, and the wandering mass fraction and wandering SMBH counts increase with $M_\star$.
\end{abstract}

%%%%%%%%%%
% Introduction
%%%%%%%%%%
\section{Introduction}
\label{sec: introduction}
Supermassive black holes (SMBHs) are thought to reside in the centers of almost all massive galaxies and have significantly shaped the evolution of the Universe. In a cosmological context, SMBHs can grow through two channels -- accretion and mergers. The accretion-driven growth primarily happens during the rapid-accretion phase when SMBHs are observed as active galactic nuclei (AGNs), and the merger-driven growth occurs after mergers of galaxies including their dark-matter halos, as predicted by the hierarchical galaxy formation model (e.g., \citealt{White78}). SMBHs follow these two growth channels to evolve from the early universe to those observed in the local universe. It is generally thought that SMBH growth is dominated by the accretion channel (e.g., \citealt{Soltan82}; \citealt{Marconi04, Kulier15, Pacucci20}) at most redshifts at which our current observational data are sensitive ($z\lesssim4$; e.g., \citealt{Zou24}), beyond which we can generally only capture the most-luminous tip of the SMBH population.\footnote{The James Webb Space Telescope (JWST) is capable of probing the SMBH growth at higher redshifts (e.g., \citealt{Maiolino24, Yang23}), but such campaigns have been advancing for only two years, and more work needs to be done before the community reaches a consensus.} However, the merger channel may become important at the high-mass/low-$z$ regime, where the available accreting material may not be sufficient to support strong accretion-driven growth (e.g., \citealt{Liu19}). Additionally, in the very early universe when SMBH seeds were formed, they may also undergo an epoch when the merger channel dominates if their number density is high (e.g., \citealt{Pacucci20, Sicilia22}), but the prediction of this phase strongly depends upon the high-redshift SMBH seeding model (e.g., \citealt{Latif16}).\par
The SMBH population, although found six decades ago with the discovery of quasars \citep{Schmidt63}, still has many unclear properties. What is their number density as a function of the black-hole mass ($M_\mathrm{BH}$) and redshift, i.e., the black-hole mass function (BHMF)? How does the scaling relation between $M_\mathrm{BH}$ and host-galaxy stellar masses ($M_\star$) evolve with redshift? How large are the contributions of the two growth channels to the overall SMBH population growth? We aim to address these questions in this work; more broadly, we seek to understand how the SMBH population evolves across cosmic time. This can be done by tracing the SMBH growth and integrating from high redshift to the local universe (e.g., \citealt{Soltan82, Marconi04}), and the accretion and merger growth channels are discussed separately below.\par
The accretion channel can be directly measured from observations. To provide a broad idea, the SMBH accretion power can be traced by the AGN luminosity, which can be converted to the black-hole accretion rate (i.e., $dM_\mathrm{BH}/dt$). When averaging over a large galaxy sample, the long-term population mean black-hole accretion rate, denoted as $\dot{M}_a$, can be measured. $\dot{M}_a$ has been shown to be closely correlated with $M_\star$ and redshift (e.g., \citealt{Yang17, Yang18, Zou24}), and thus we can use $(M_\star, z)$ as predictors of $\dot{M}_a$ for the galaxy population. This is important because $M_\star$ and $z$ are the most straightforward physical parameters to measure with modern multiwavelength photometric surveys (e.g., \citealt{Zou22}). In \citet{Zou24}, we provided the best current measurement of $\dot{M}_a$ as a function of $(M_\star, z)$ based on a large compilation of data from nine high-quality survey fields (spanning $0.05-60~\mathrm{deg^2}$ following a wedding-cake design) and a semiparametric Bayesian method. \citet{Zou24} adopted the \mbox{X-ray} emission as the accretion-power tracer, which is universal among AGNs, has high penetrating power through obscuration, and suffers little from starlight dilution (e.g., \citealt{Brandt15, Brandt22, Xue17}). All of these factors enabled reliable measurements of $\dot{M}_a$ across a wide parameter space, covering $9.5<\log(M_\star/M_\odot)<12$ and redshifts up to 4. \citet{Zou24} also showed that this $\dot{M}_a(M_\star, z)$ function, after convolution with the galaxy stellar mass function (SMF), can successfully return an AGN \mbox{X-ray} luminosity function consistent with direct observations.\par
The merger channel, however, can hardly be directly constrained by observations, at least for now. In principle, SMBH mergers can be probed through nanohertz gravitational waves (GWs) observed by pulsar timing arrays (e.g., see \citealt{Mingarelli19} for a brief review), and recent results have indeed begun to provide promising insights (e.g., \citealt{Agazie23}). However, the currently observed nanohertz GW background still has a low signal-to-noise ratio and cannot provide effective constraints on SMBH mergers. Fortunately, the merger channel is accessible via cosmological simulations. Modern cosmological simulations are now able to self-consistently and simultaneously reveal both large structures up to hundreds of Mpc and internal physics within individual galaxies. Galaxies merge during the simulated cosmic evolution, which may lead to SMBH mergers, but the eventual SMBH coalescence status depends upon whether there is sufficient time for the SMBHs to dissipate their orbital energy. Unlike accretion, which is directly linked to the vicinity of SMBHs, the overall galaxy merger history is largely set by the much larger environment at the dark-matter halo scale. Therefore, merger histories are generally more reliable in simulations compared to simulated accretion. We rely on the IllustrisTNG project\footnote{\url{https://www.tng-project.org}} \citep{Weinberger17, Marinacci18, Naiman18, Nelson18, Nelson19a, Nelson19b, Pillepich18a, Pillepich18b, Pillepich19, Springel18}, a series of gravo-magnetohydrodynamical cosmological simulations, to extract the merger information.\par
In this work, we use a hybrid method to combine the accretion channel from observations and the merger channel from simulations, as discussed above, to integrate over time and predict the overall cosmic evolution of the SMBH population. There have been valuable efforts to understand the SMBH population evolution from either the observational or the theoretical sides, although the exact boundaries between \textit{theories} and \textit{observations} are sometimes vague. Indeed, the SMBH population is widely implemented in semi-analytic models (SAMs) and hydrodynamical simulations for galaxy evolution (e.g., \citealt{Kulier15, Sijacki15, Somerville15, Weinberger17, Ricarte18}). However, these theoretical works inevitably have to simplify the complex physics governing the SMBH and galaxy evolution, which may cause challenges in having realistic characterization of SMBHs. For example, IllustrisTNG and many other cosmological simulations have already included the SMBH population, which means that, in principle, they can model the SMBH accretion. However, given the limited resolution, these simulated SMBHs are primarily modeled with largely uncertain sub-grid physics. It should also be noted that these large-scale cosmological simulations are not calibrated against the observed AGN population such as the AGN luminosity function or $\dot{M}_a(M_\star, z)$. Even for detailed, resolved modeling focusing on the AGN disk and feedback, simulations still have significant inconsistencies compared with observations (e.g., \citealt{Lawrence18, Davis20}; and references therein). It has been shown that the SMBH population in cosmological simulations cannot match well with observations (e.g., \citealt{Aird18, Habouzit21, Habouzit22}), and we will also illustrate this point in our following sections. As a complementary approach, observational constraints generally do not require assumptions about the underlying physics, but instead connect the $M_\mathrm{BH}$ growth to accretion-driven electromagnetic radiation or host-galaxy properties (e.g., \citealt{Soltan82, Marconi04, Shankar09, Yang18}). Some models also combine observations and theories to characterize SMBHs. For example, \citet{Ricarte18} implemented the observation-motivated accretion channel in their SAM, \citet{Shankar20a} combined the observed accretion channel and the SAM-predicted galaxy evolutionary tracks, and \citet{Zhang23} presented a data-driven empirical model constrained by various observed results. These approaches generally show more realistic modeling of the SMBH and AGN population than purely theoretical results. However, such observational constraints usually lack direct, careful treatments of mergers. Due to these reasons, we aim to take advantage of both observations and simulations to have a more realistic depiction of SMBH growth. Instead of focusing on setting appropriate models for the SMBH accretion to match observations, we will directly implement the observed accretion channel; when combined with simulations, we can further derive the contribution from the merger channel. Besides, this work benefits from both the state-of-the-art observational and simulation results that became available only within recent years. Our adopted new results in \citet{Zou24} significantly improve the constraints for the accretion channel, and the IllustrisTNG simulations have been intensively explored and proven to be successful over the past six years.\par
This work is structured as follows. Section~\ref{sec: methodology} describes the methodology. Section~\ref{sec: results} presents our results and discusses their implications. Section~\ref{sec: summary} summarizes this work. We adopt a flat $\Lambda\mathrm{CDM}$ cosmology with $H_0=70~\mathrm{km~s^{-1}~Mpc^{-1}}$, $\Omega_\Lambda=0.7$, and $\Omega_M=0.3$.

%%%%%%%%%%
% Methodology
%%%%%%%%%%
\section{Methodology}
\label{sec: methodology}
\subsection{Overview}
\label{sec: method_overview}
Briefly, to reconstruct the SMBH population at different redshifts, we begin at $z=4$ with some given initial conditions and let SMBHs grow through both the accretion and merger channels to evolve to $z=0$. We only probe redshift up to 4 because we do not have reliable observational constraints at higher redshift (e.g., \citealt{Zou24}).\par
The TNG simulations provide properties including $M_\star$ of each galaxy as a function of $z$ and their merger trees. Starting at $z=4$, we seed SMBHs\footnote{Note that the term \textit{seed} in our context is irrelevant to the earliest SMBH seeds at even higher redshift. We only mean that we set $M_\mathrm{BH}$ for massive galaxies as initial conditions.} into the TNG-simulated galaxies that have $M_\star>10^{9.5}~M_\odot$ for the first time using the $M_\mathrm{BH}-M_\star$ scaling relation calibrated against broad-line AGNs in \citet{Reines15}, $\log(M_\mathrm{BH}/M_\odot)=7.45+1.05[\log(M_\star/M_\odot)-11]$, as will be discussed in Section~\ref{sec: bhseed}. We add a typical scatter of 0.5~dex to the seeded $M_\mathrm{BH}$ (e.g., \citealt{Li23, Reines15}). This means that, at $z=4$, we seed all the galaxies with $M_\star(z=4)>10^{9.5}~M_\odot$ while setting the $M_\mathrm{BH}$ of the remaining less-massive galaxies to be 0; once a low-mass galaxy accumulates sufficient $M_\star$ to reach $10^{9.5}~M_\odot$ at lower redshift, we then seed a SMBH into it. Therefore, considering the main progenitors, SMBHs are not necessarily seeded at $z=4$. The bottom left panel of Figure~\ref{fig: mbhevo} will present typical $M_\star$ evolutionary tracks. Briefly, galaxies with $M_\star(z=0)>10^{10.5}~M_\odot$ generally reach $10^{9.5}~M_\odot$ at $z\gtrsim3$ and are thus seeded sufficiently early such that the majority of their SMBH growth is captured. However, low-mass galaxies with $M_\star(z=0)<10^{10}~M_\odot$ reach $10^{9.5}~M_\odot$ much later ($z\lesssim1$), and hence a considerable fraction of their SMBH growth may be missed in our analyses, leading our results in the low-mass regime to be more dominated by the choice of seeding.\par
We let the SMBHs grow with a rate predicted by the median $\dot{M}_a(M_\star, z)$ function in Section~3.3 of \citet{Zou24}. When two galaxies merge according to the TNG merger trees, we either let the two SMBHs (if any) in these galaxies merge or only keep the more massive one in the newly merged galaxy, as will be detailed in Section~\ref{sec: merger}. For a given galaxy labeled with a subscript of $i$ evolving from $z$ to $z-\delta z$, the above procedures can be summarized as the following equation,
\begin{align}
M_{\mathrm{BH}, i}(z-\delta z)=&M_{\mathrm{BH}, i}(z)+\dot{M}_a(M_{\star, i}(z), z)\delta t\nonumber\\
&+M_{\mathrm{BH}, i}^\mathrm{merger}(z\to z-\delta z),
\end{align}
where $\delta t$ is the cosmic time corresponding to $\delta z$, and $M_{\mathrm{BH}, i}^\mathrm{merger}(z\to z-\delta z)$ is the increase in $M_{\mathrm{BH}, i}$ from $z$ to $z-\delta z$ through mergers. There are 79 TNG snapshots at $z\leq4$ with a time step of $\approx0.15$~Gyr. We iteratively follow these procedures until $z=0$ to recover the expected SMBH population between $0\leq z\leq4$. This approach of integrating the accretion across time is broadly related to the classical Soltan argument \citep{Soltan82}, which has been written in differential forms (e.g., \citealt{Small92, Marconi04}) and also modified in \citet{Shankar20a} to integrate along $M_\star$ evolutionary tracks. There are several important points worth noting regarding these procedures, and we will discuss them in the subsequent subsections.

\subsection{SMBH Seeds}
\label{sec: bhseed}
We only seed SMBHs into galaxies more massive than $10^{9.5}~M_\odot$, which means that we do not attempt to probe less-massive galaxies. This is because the massive black holes (MBHs)\footnote{The term \textit{SMBH} is usually not used for these dwarf galaxies because their hosted black holes may not be sufficiently massive to be SMBHs but may instead be intermediate-mass black holes. We thus use the general term \textit{MBH} here.} in dwarf galaxies with $M_\star<10^{9.5}~M_\odot$ are still poorly understood, and we currently do not have a well-measured $\dot{M}_a(M_\star, z)$ function in this low-mass regime (e.g., \citealt{Zou23, Zou24}). In fact, dwarf galaxies do not always necessarily host MBHs, and observational constraints on their MBH-occupation fraction are still loose (e.g., \citealt{Miller15, Gallo19}). The accretion-driven growth of MBHs in dwarf galaxies also does not necessarily mainly happen through typical AGN gas accretion but may be substantially powered by tidal-disruption events (e.g., \citealt{Zubovas19}).\par
Our seeding follows the $M_\mathrm{BH}-M_\star$ relation calibrated against broad-line AGNs in \citet{Reines15}. They reported separate scaling relations for AGNs and inactive galaxies with dynamical $M_\mathrm{BH}$ measurements and showed that these two relations can differ by over one order of magnitude. \citet{Shankar16, Shankar20a, Shankar20b} argued that the AGN one is more accurate because the relation based on dynamically measured $M_\mathrm{BH}$ may be significantly biased towards those with more massive SMBHs whose spheres of influence are easier to resolve. Our results later (Section~\ref{sec: scaling}) also seem to be more consistent with the AGN-based relation. We will discuss the impacts of the adopted seeding relation on our results in Section~\ref{sec: differentseed}.\par
Furthermore, by using the local-universe $M_\mathrm{BH}-M_\star$ relation at all redshifts, we implicitly adopt the assumption that this scaling relation does not evolve with redshift, at least up to $z=4$. This is a strong assumption that needs careful assessment. It has been shown in recent works that this scaling relation does not show a significant redshift evolution (e.g., \citealt{Yang18, Yang19, Yang22, Shankar20a, Suh20, Sun15, Habouzit21, Li21a, Li21b, Li23}), though direct observational assessments can hardly reach $z>1$. Our results in Section~\ref{sec: scaling} will show that our predicted scaling relation indeed does not have a large variation across different redshifts. One side point worth noting is that, at higher redshifts above $z=4$, there is some emerging evidence suggesting that SMBHs may be overmassive (e.g., \citealt{Pacucci24}), but it is still under debate whether the high-redshift $M_\mathrm{BH}-M_\star$ relation is indeed higher than the local-universe one, and the corresponding possible evidence is mainly from the early universe, which is beyond the scope of this work and would not materially affect our conclusions. We will also try a redshift-dependent seeding relation in Section~\ref{sec: differentseed}.\par
Furthermore, we adopt a single typical seeding scatter of 0.5~dex for all the redshifts. It is unclear how large the $M_\mathrm{BH}-M_\star$ scatter is at higher redshifts, and different simulations return different results \citep{Habouzit21}. Nevertheless, we expect the seeding scatter to contribute little to our overall $M_\mathrm{BH}$ scatter at $M_\star\gtrsim10^{10}~M_\odot$ and $z<3$, as explored in Section~\ref{sec: scaling}.

\subsection{The Accretion Channel}
\label{sec: accretion}
Our chosen growth rate from Section~3.3 of \citet{Zou24} only depends on $(M_\star, z)$ but not on the SMBH itself or its vicinity. This approach has an important advantage in that $M_\star$, as a global parameter of a galaxy, is generally more reliable to measure than local parameters, and utilizing the observational constraints in \citet{Zou24} avoids relying on oftentimes largely uncertain sub-grid physics to depict the SMBH accretion, as done in simulations.\par
The $\dot{M}_a$ in \citet{Zou24} is the mean $M_\mathrm{BH}$ growth rate averaged over all the galaxies with given $(M_\star, z)$. The short-term AGN phase has been averaged out during this process, and thus $\dot{M}_a$ reflects the long-term population mean. \mbox{X-rays} were used to sample the accretion power, and there are two additional quantities assumed. One is the bolometric correction converting \mbox{X-ray} luminosities to bolometric luminosities, and the luminosity-dependent bolometric correction in \citet{Duras20} was adopted. The other is the accretion radiative efficiency converting bolometric luminosities to mass-accretion rates, which is assumed to be a canonical value of 0.1 (e.g., \citealt{Davis11}). These conversions have typical systematic uncertainties up to a factor of a few. An additional systematic uncertainty arises from the fact that the \mbox{X-ray}-constrained $\dot{M}_a$ in \citet{Zou24} may miss Compton-thick accretion. The typical fraction of Compton-thick AGNs at $z=0$ is generally constrained to be $\approx20\%-50\%$ (e.g., \citealt{Ricci15, Kammoun20}; Boorman et al., submitted). Using the \mbox{X-ray} luminosity function with the column-density distribution included in Sections 3 and 6 of \citet{Ueda14}, we found that the fractional accretion power at column densities above $10^{24}~\mathrm{cm^{-2}}$ is $\approx38\%$ for all the redshifts. Besides, some Compton-thick AGNs are still able to be detected (e.g., \citealt{Li19, Yan23}), especially given the fact that we are probing higher rest-frame energies (reaching $15-40$~keV) with higher penetrating power as redshift increases. Therefore, the systematic bias of $\dot{M}_a$ from possibly missed Compton-thick accretion should be $\lesssim0.2$~dex, where the upper limit, 0.2~dex, corresponds to the case where all the Compton-thick accretion accounting for 38\% of the total accretion power is missed. One caveat worth noting is that there is emerging evidence suggesting that more accretion power may be hidden by heavy obscuration at high redshift than our previous expectations. For example, \citet{Yang21} showed that the \mbox{X-ray}-inferred $\dot{M}_a$ may be underestimated by a factor of a few at $z>3$, and a large number of \textit{little red dots} at $z\gtrsim4$ were recently found by JWST and might contain heavily obscured AGNs (e.g., \citealt{Kocevski24}). However, this possible discrepancy seems to mainly exist at high redshift and barely impacts the redshift range on which we are focusing, where most cosmic SMBH growth occurs.\par
Besides the dependence on $(M_\star, z)$, $\dot{M}_a$ also depends on star formation rate (e.g., \citealt{Yang19}), galaxy morphology (e.g., \citealt{Ni21}), and perhaps cosmic environment (e.g., \citealt{Song16, Yang18env}). However, for the general galaxy population, the dependence on star formation rate and environment is only secondary compared to the $M_\star$ dependence (e.g., \citealt{Yang17, Yang18env, Yang19, Amiri19}), and the morphological dependence, though possibly more fundamental, is still not fully understood in a quantitative sense because morphological measurements are expensive to obtain. Therefore, $(M_\star, z)$ should still be the best available predictors for $\dot{M}_a$ that capture most of the variance.\par
There are statistical uncertainties when measuring $\dot{M}_a(M_\star, z)$ in \citet{Zou24}, which are not included in our analyses. Appendix~\ref{append: adderr} shows that these would only add a scatter of $\approx0.05-0.1$~dex into our $M_\mathrm{BH}$ and are thus negligible compared to the aforementioned systematic uncertainties.

\subsection{SMBH Mergers}
\label{sec: merger}
Regarding the SMBHs after two galaxies merge, they usually undergo a long process before they finally reach coalescence, as discussed in, for example, \citet{Yu02}, \citet{Barausse21}, and references therein. The two SMBHs will first wander independently in the merged galaxy with a 10-kpc scale separation and gradually lose their orbital energy through the Chandrasekhar dynamical friction before sinking into the galaxy center and forming a close SMBH pair with a sub-kpc separation (e.g., \citealt{Dosopoulou17}). The SMBHs then harden at a pc-scale separation where the dynamical friction process becomes inefficient, and the orbital energy will be dissipated through dynamical interactions with stars (e.g., \citealt{Quinlan96}), gas (e.g., \citealt{Lodato09}), or even another SMBH (e.g., \citealt{Bonetti19}). Once the binary separation shrinks to $\approx10^{-3}-10^{-2}~\mathrm{pc}$, the orbit will rapidly decay through GW emission and finally lead to a SMBH merger event. The whole process may take longer than the Hubble time, in which case the SMBH merger will be stalled (e.g., \citealt{Dosopoulou17, Tremmel18, Ricarte21a, Ricarte21b}), and thus SMBHs only have a less-than-unity probability of eventually merging ($p_\mathrm{merge}$) after their host galaxies merge. Those that do not merge will constitute a long-lived \textit{wandering} SMBH population.\par
\begin{figure}
\includegraphics[width=\hsize]{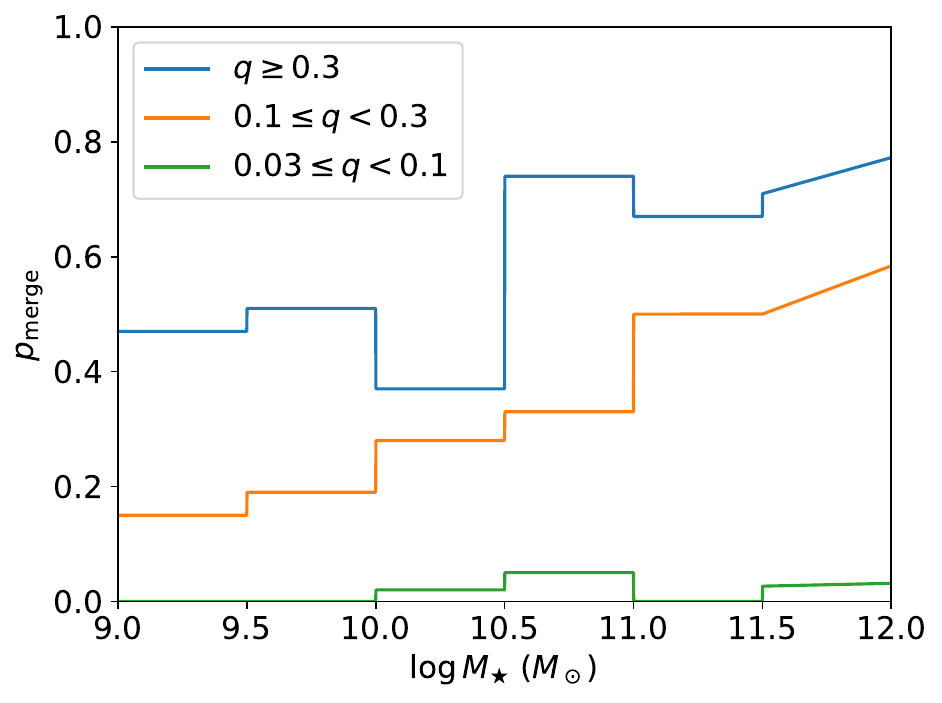}
\caption{Our adopted $p_\mathrm{merge}$ based on \citet{Tremmel18} as a function of $M_\star$ at different $q$ bins. $p_\mathrm{merge}$ is set to 0 for $q<0.03$. $p_\mathrm{merge}$ is extrapolated only in the massive end ($M_\star>10^{11.5}~M_\odot$).}
\label{fig: pmerge}
\end{figure}

We adopt $p_\mathrm{merge}$ in Figure~1 of \citet{Tremmel18}, who presented this quantity as a function of the $M_\star$ ratio $q$ ($\leq1$) of the two merging galaxies and the $M_\star$ of the primary galaxy within $q>0.03$ and $10^{9}<M_\star<10^{11.5}~M_\odot$. Such an approach was also adopted in \citet{Pacucci20}. The $p_\mathrm{merge}$ relation in \citet{Tremmel18} is derived from simulations with emphasis on modeling SMBHs such as their dynamics in the context of cosmological simulations and should be one of the best ones currently available. $p_\mathrm{merge}$ strongly increases with $q$ and $M_\star$, where $p_\mathrm{merge}\approx0$ for minor mergers with $0.03<q<0.1$, and reaches $\approx0.7$ for major mergers in massive galaxies with $M_\star>10^{10.5}~M_\odot$. Therefore, we set $p_\mathrm{merge}=0$ for $q<0.03$, which is not covered in \citet{Tremmel18}; we also use a linear relation to fit the $p_\mathrm{merge}-\log M_\star$ relation at a given $q$ bin in \citet{Tremmel18} and extrapolate it into $M_\star>10^{11.5}~M_\odot$. We show $p_\mathrm{merge}$ versus $M_\star$ in Figure~\ref{fig: pmerge}. In principle, $p_\mathrm{merge}$ also strongly depends upon the galaxy structure such as the central stellar density and spherical asymmetry (e.g., \citealt{Vasiliev14, Tremmel18}), but it is difficult to directly apply these relations, and we reiterate that our analyses only aim to work in an average sense. Besides, \citet{Tremmel18} only accounted for the SMBH-binary evolution down to a separation of $\approx700$~pc. This indicates that the resulting $p_\mathrm{merge}$ is only an upper limit, although the SMBH-binary evolution time is generally expected to be relatively shorter than the evolution time before reaching this separation. We note that the evolution of SMBH binaries from a galactic scale down to a final coalescence is still under active research and beyond the scope of this work, and the merger rate heavily depends on the complex physics briefly outlined above, as extensively explored in, for example, \citet{Barausse20}. It often cannot be fully modeled even at a galactic scale in large cosmological simulations; for example, the TNG simulations themselves adopt a simple, so-called \textit{repositioning} strategy to manually place the SMBH positions into the minimum of the gravitational potential at each time step \citep{Weinberger17}, which would cause a much overestimated SMBH merger rate \citep{Chen22}. Here, we turn to reasonable simplified approximations to quantify $p_\mathrm{merge}$ and will also explore the sensitivity of our results on the choice of $p_\mathrm{merge}$. Especially, we will try extreme cases of $p_\mathrm{merge}=0$ and $p_\mathrm{merge}=1$ in Section~\ref{sec: differentpmerge}. We will still focus on results based on $p_\mathrm{merge}$ in \citet{Tremmel18} in the main text because it is more physically meaningful and contains expected dependencies on $q$ and $M_\star$.\par

When two SMBHs eventually merge, they will also lose mass due to GW radiation. The remnant $M_\mathrm{BH}$ depends on the SMBH properties in a complex manner, and we assume that, typically, 10\% of the less-massive $M_\mathrm{BH}$ is lost due to GW radiation (e.g., \citealt{Healy14, Varma19}). Such a mass loss is only secondary compared to the merger probability and its large uncertainty. It is also worth noting that the GW radiation has recoil upon merging SMBHs, which might cause some SMBHs to be kicked out of the system (e.g., \citealt{Lousto12}). However, this can hardly affect the overall $p_\mathrm{merge}$ as only $\lesssim5\%$ of SMBH binaries in the GW-radiation phase are expected to be ejected (e.g., \citealt{Lousto12, Kulier15}), but it is possible that the GW-recoiled wandering SMBHs may be dominant in dark-matter halo centers \citep{IzquierdoVillalba20}.\par
To summarize, if two galaxies, both of which contain central SMBHs, merge according to the merger trees, we first randomly determine if the two SMBHs will merge with a single Bernoulli realization with a success probability of $p_\mathrm{merge}$. If a merger is expected, we let it happen immediately and generate a single SMBH with the GW-radiation mass loss subtracted. If a merger does not happen, we keep the more massive SMBH as the central one that will continue to grow in future snapshots and put the other one into the wandering SMBH population that will not grow later, and our assumption that wandering SMBHs do not grow has been attested to be generally appropriate in \citet{Ricarte21b}. We acknowledge that there is generally a time delay of several hundred Myrs between galaxy mergers and the final SMBH coalescence (e.g., \citealt{Tremmel18}). However, such an effect strongly depends on the physical processes discussed above and is difficult to quantify. It is also challenging to adopt a single value to represent the typical delay time because \citet{Tremmel18} showed that the delay time is widely distributed, with half of the merged SMBHs having small time delays ($<0.5$~Gyr) and the other half distributed in a heavy tail of the delayed time distribution extending to several Gyrs. As far as we know, there are actually no well-constructed relations available between the delay time and global galaxy properties. Besides, our TNG snapshots have a time step of 150~Myrs, and thus the typical delay timescale corresponds to only a few time steps and is negligible to the overall cosmic time span. Therefore, we do not add any time delays here to the SMBH merger for simplicity. The systematic error caused by this simplification should be smaller than that in the extreme case of $p_\mathrm{merge}=0$, which, as we will show in Section~\ref{sec: differentpmerge}, does not have major effects upon most of our results. Besides, a time delay does not change the overall merged mass but only smears out the mergers over time, and thus it would not cause any difference in population properties if the population mean merger-driven $M_\mathrm{BH}$ growth rate ($\dot{M}_m$) is independent of redshift. As Section~\ref{sec: channelgrow} will show, $\dot{M}_m$ only weakly correlates with redshift for all the $M_\mathrm{BH}$ or $M_\star$ values, and the changes in $\dot{M}_m$ within 1~Gyr are $\lesssim0.2$~dex; thus delayed mergers will not affect $\dot{M}_m$ by over 0.2~dex.

\subsection{TNG}
TNG includes ten baryonic simulations in total on three choices of box side lengths, and the flagship, highest-resolution versions of each box side length choice include TNG50-1, TNG100-1, and TNG300-1; see \citet{Nelson19b} for more details. We do not explicitly write the suffix ``1'' when referring to these flagship TNG simulations hereafter for conciseness. Briefly, these simulations share the same baryonic physics model; side lengths of $35~h^{-1}$, $75~h^{-1}$, and $205~h^{-1}$ (i.e., roughly 50, 100, and 300 for $h=H_0/(100~\mathrm{km~s^{-1}~Mpc^{-1}})=0.7$) comoving Mpc are used for TNG50, TNG100, and TNG300, respectively, while their baryonic mass resolutions are roughly $10^5$, $10^6$, and $10^7~M_\odot$, respectively.\par
As discussed in Section~\ref{sec: method_overview}, we do not directly utilize the SMBH population in TNG. However, we briefly summarize the TNG SMBH modelling for comparison with our own accretion channel. Full details of the TNG SMBHs were described in \citet{Weinberger17, Weinberger18}. TNG SMBHs are seeded with $M_\mathrm{BH}=1.2\times10^6~M_\odot$ when dark-matter halos reach $7.4\times10^{10}~M_\odot$. Their accretion rates are assumed to follow the Bondi-Hoyle-Lyttleton model (\citealt{Edgar04} and references therein), which is determined by the ambient gas properties, with an upper limit of the Eddington accretion rate. SMBHs further cause feedback in two modes -- kinetic and thermal energy is released into the environment when the accretion rates are low and high, respectively. The transition accretion rate between these two modes depends on $M_\mathrm{BH}$ such that more massive SMBHs are more likely to be in the kinetic mode. Regarding the SMBH dynamics, TNG repositions SMBHs in the local potential minimum at each global time step and merges SMBH binaries without a time delay. The simulation parameters are tuned such that the predicted $M_\mathrm{BH}-M_\star$ relation at $z=0$ broadly matches the observed, dynamically measured relation, and several other predicted basic galaxy-population properties can match observations with the SMBH feedback (see \citealt{Weinberger17} and \citealt{Pillepich18b} for more details).\par
It has been shown that the simulated galaxy population is generally in good agreement with observations (e.g., \citealt{Pillepich18b}), such as their SMFs \citep{Pillepich18a}, colors \citep{Nelson18}, and morphology \citep{Rodriguez-Gomez19}. Nevertheless, the consistency, although remarkable, is not perfect. Small differences of $\lesssim0.3$~dex exist in terms of the SMF, and the deviations can reach over 1~dex at the very massive end ($M_\star\gtrsim10^{12}~M_\odot$), depending upon the $M_\star$ definition in simulations \citep{Pillepich18a}. These values reflect the typical systematic uncertainty of our results originating from simulations. The cosmic variance within finite simulation boxes generally introduce smaller uncertainties. \citet{Genel14} showed that the cosmic variance only introduces an uncertainty of $\lesssim0.1$~dex to the SMF at $z=0$ for a TNG100 box, and the uncertainty only increases to 0.3~dex in the massive end of the SMF at $z=3$. This cosmic variance will be further suppressed and become negligible for a TNG300 box.\par
We conduct our analyses for TNG100 and TNG300, both of which have sufficiently large galaxy source statistics. We do not analyze TNG50 for two reasons. First, its volume is small and only contains less than 300 galaxies with $M_\star>10^{10.5}~M_\odot$ at $z=0$. Secondly, it is not necessarily more reliable than TNG100 in terms of the $M_\star$ evolution of massive galaxies. TNG50 was designed for a different purpose of probing the structural and kinematic properties down to a $\approx0.1$-kpc scale \citep{Nelson19a, Pillepich19}, and it is TNG100, not TNG50, that is calibrated against the observed galaxy population. Therefore, TNG50 cannot provide more insights into our specific science than for TNG100.\par
We adopt the \texttt{SubLink} TNG merger trees derived based on the algorithm in \citet{Rodriguez-Gomez15}. We adopt $M_\star$ as the sum of stellar particles within twice the stellar half mass radius, i.e., \texttt{SubhaloMassInRadType[4]} in the TNG catalogs. It is known that simulated galactic properties generally depend on the resolution. For $M_\star$, it generally increases with better resolution, as explored in, e.g., \citet{Pillepich18a, Pillepich18b}. To mitigate this effect, we rescale the TNG300 $M_\star$ so that the TNG300 SMF at $M_\star>10^{9.5}~M_\odot$ can match the best with the TGN100 one. We use TNG100 as the reference because it was calibrated with observations, while other TNG simulations are not directly calibrated but share the same physics with TNG100. Following the suggestion in \citet{Pillepich18a}, we set the rescaling factor to be redshift dependent but not $M_\star$ dependent, which is sufficient for a good correction for the resolution effect, and the typical value is 1.5. Therefore, the resolution rescaling correction is at most 0.2~dex, consistent with the finding in \citet{Pillepich18a}.\par
We select and only analyze galaxies with $M_\star>10^{9.5}~M_\odot$ at $z=0$ and their progenitors in the merger trees because, as has been explained in Section~\ref{sec: bhseed}, our understanding of the MBH accretion in dwarf galaxies is still poor. Galaxies not included in merger trees are not considered; such cases are very rare for galaxies with masses far above the mass-resolution limits, and only 1\% of galaxies with $M_\star>10^{9.5}~M_\odot$ at $z=0$ are excluded. This returns us 10635 and 186542 galaxies at $z=0$ for TNG100 and TNG300, respectively. The numbers of their progenitors with $M_\star>10^{9.5}~M_\odot$ at $z=1$, 2, 3, and 4 are 8615 (156018), 5393 (101663), 2553 (49630), and 970 (19355) for TNG100 (TNG300), respectively. To keep the narrative flow concise, we focus on TNG300 in the main text because it has better number statistics than TNG100. We will not explicitly write ``TNG300'' hereafter, and unless noted in the main text, we always refer to TNG300. TNG100 and TNG300 results are generally similar, as illustrated in Appendix~\ref{append: comptng}.

%%%%%%%%%%
% Results
%%%%%%%%%%
\section{Results}
\label{sec: results}
\begin{figure*}
\resizebox{\hsize}{!}{
\includegraphics{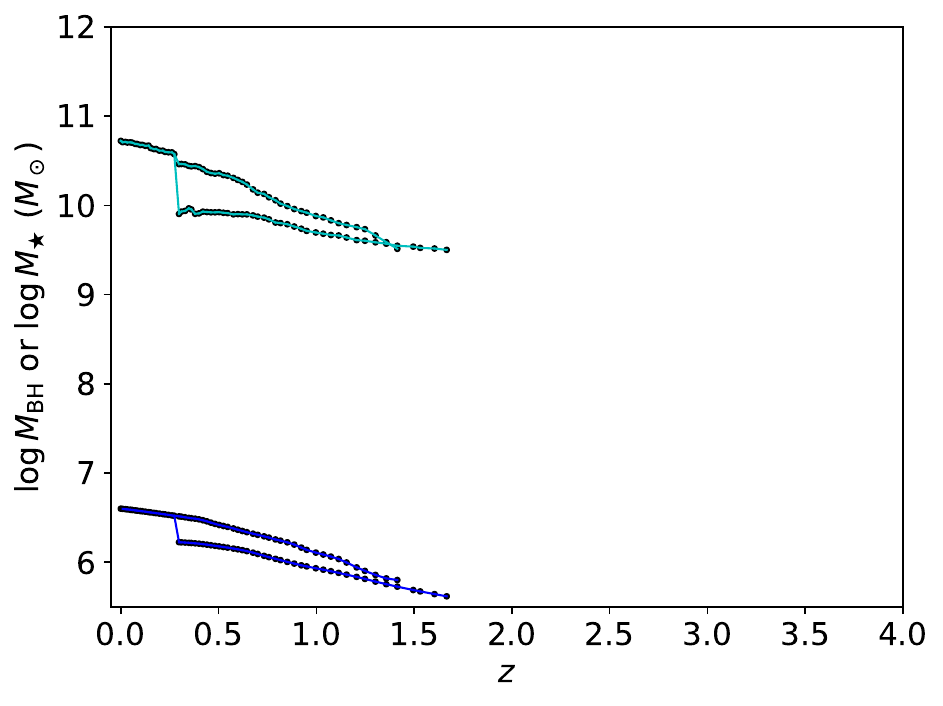}
\includegraphics{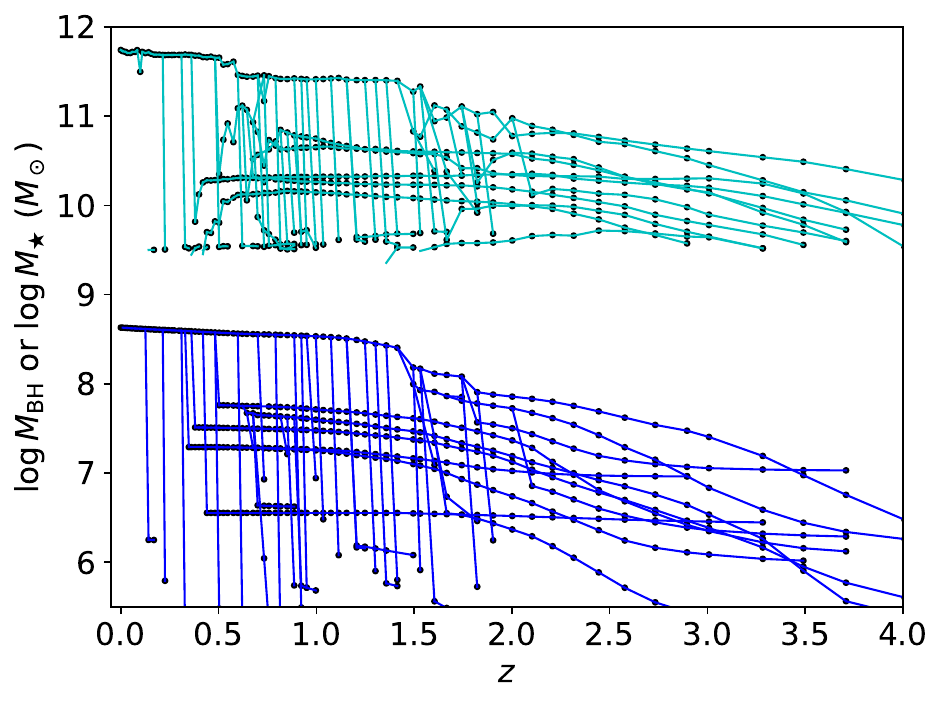}
}
\caption{Example $M_\star$ and $M_\mathrm{BH}$ evolutionary tracks for MW-like galaxies (left) and massive galaxies with $M_\star\approx10M_\star^\mathrm{MW}$ (right). Each panel presents one galaxy at $z=0$ and all of its progenitors with $M_\star>10^{9.5}~M_\odot$ at previous snapshots. Each black point marks one massive galaxy with $M_\star>10^{9.5}~M_\odot$ at the corresponding snapshot, and dwarf galaxies are not shown for visual clarity. Cyan and blue lines connect galaxies with their descendants in terms of $M_\star$ and $M_\mathrm{BH}$, respectively. A few cyan lines drop down at $\approx10^{9.5}~M_\odot$ without connecting to descendant points because their descendants have $M_\star<10^{9.5}~M_\odot$ and are thus not shown, but these galaxies will eventually merge to form the final $z=0$ galaxy. The massive-galaxy example in the right panel apparently undergoes more mergers with massive galaxies than for the MW-like example in the left panel, which undergoes only one merger with massive galaxies.}
\label{fig: evotracks}
\end{figure*}

\begin{figure}
\includegraphics[width=\hsize]{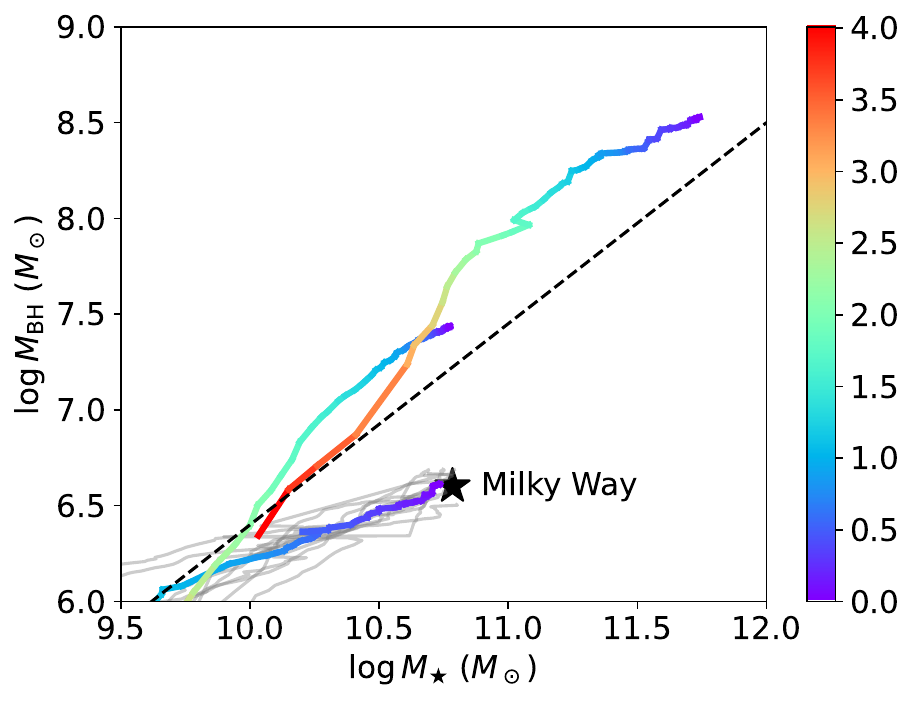}
\caption{Example evolutionary tracks on the $M_\mathrm{BH}-M_\star$ plane. The black star marks the MW. The gray curves represent individual MW-like galaxies. The other three multicolor curves are color-coded by the redshift. The one connected to the black star is the median track of MW-like galaxies, and the other two are the median tracks of galaxies with $M_\star\approx M_\star^\mathrm{MW}$ and $M_\star\approx10M_\star^\mathrm{MW}$. The dashed line is the AGN-based $M_\mathrm{BH}-M_\star$ relation in \citet{Reines15}.}
\label{fig: examplescalingevo}
\end{figure}

Before analyzing the overall population, we show typical example evolution results for galaxies similar to the Milky Way (MW) and more massive ones to provide first insights. The MW has $M_\star^\mathrm{MW}=6.08\times10^{10}~M_\odot$ (e.g., \citealt{Licquia15}) and $M_\mathrm{BH}^\mathrm{MW}=4.0\times10^6~M_\odot$ (e.g., \citealt{EHT22}), and we define MW-like galaxies as those within $\log M_\star^\mathrm{MW}\pm0.1$~dex and $\log M_\mathrm{BH}^\mathrm{MW}\pm0.1~\mathrm{dex}$ at $z=0$. We also select massive galaxies with $M_\star$ higher than $M_\star^\mathrm{MW}$ by $0.9-1.1$~dex. Figure~\ref{fig: evotracks} presents examples of the $M_\star$ and $M_\mathrm{BH}$ evolution for MW-like galaxies and massive ones. MW-like galaxies typically reach our $M_\star$ threshold of $10^{9.5}~M_\odot$ at intermediate redshifts of $\approx1.5-2$ and undergo few mergers with massive galaxies, while the massive ones are already above the $M_\star$ threshold at $z=4$ and undergo many mergers during their evolution. Figure~\ref{fig: examplescalingevo} further shows their evolutionary tracks in the $M_\mathrm{BH}-M_\star$ plane. The MW hosts a smaller SMBH compared to the median $M_\mathrm{BH}$ of other galaxies with similar $M_\star$ by a factor of 7 (0.8~dex), and thus the tracks of our MW-like galaxies are generally below the $M_\mathrm{BH}-M_\star$ relation in \citet{Reines15}. Compared to other galaxies with similar $M_\star$, the $M_\star$ growth of these MW-like galaxies happens at lower redshifts. This is required to produce a smaller $M_\mathrm{BH}$ in our methodology. Since $\dot{M}_a$ increases with rising redshift at a given $M_\star$, the later $M_\star$ growth of MW-like galaxies can lead to smaller $\dot{M}_a$. We note that the MW does not stand out as a strong outlier because the $M_\mathrm{BH}$ scatter of our sources with $M_\star\approx M_\star^\mathrm{MW}$ is 0.4~dex due to the diverse growth history of $M_\star$, and this scatter is only a lower limit because the $(M_\star, z)$ predictor set does not capture all the variance of $\dot{M}_a$ (Section~\ref{sec: scaling}). We do not attempt to link the mass assemblies of these MW-like galaxies to the mass assembly of the actual MW, which has been an open question for decades. For example, we compared the estimated MW mass-assembly histories in different literature works (e.g., \citealt{Fantin19, Xiang22}) with those of MW-like galaxies in \citet{vanDokkum13} and did not find consistent evidence suggesting the MW mass assembly is relatively earlier or later. The MW SMBH may also have undergone a previous major merger (e.g., \citealt{Wang24}) that further complicates the growth history. Instead, we solely use the examples above to illustrate how a typical SMBH evolves together with its host galaxy and how different $M_\star$ evolutionary tracks can lead to different $M_\mathrm{BH}$.

We will now present the overall population analyses in the following subsections. Specifically, Section~\ref{sec: scaling} presents the $M_\mathrm{BH}-M_\star$ scaling relation and its cosmic evolution; Section~\ref{sec: bhmf} presents the BHMF and its cosmic evolution; Section~\ref{sec: channelgrow} analyzes the growth rates in the accretion and merger channels; Section~\ref{sec: massassembly} presents the overall redshift evolution of $M_\mathrm{BH}$, Section~\ref{sec: wander} discusses wandering SMBHs; Section~\ref{sec: differentpmerge} discusses the impacts of the $p_\mathrm{merge}$ choice on our results; and Section~\ref{sec: differentseed} discusses the impacts of the SMBH seeding choice on our results.

% Scaling relation
\subsection{The $M_\mathrm{BH}-M_\star$ Relation}
\label{sec: scaling}
$M_\mathrm{BH}$ has been shown to follow a correlation with the host bulge mass and also $M_\star$, which is an indication of the coevolution between SMBHs and galaxies (e.g., \citealt{Kormendy13} and references therein). Although the correlation with the bulge is tighter, we focus on $M_\star$ because our whole methodology is integrated based on $M_\star$.\par
\begin{figure}
\includegraphics[width=\hsize]{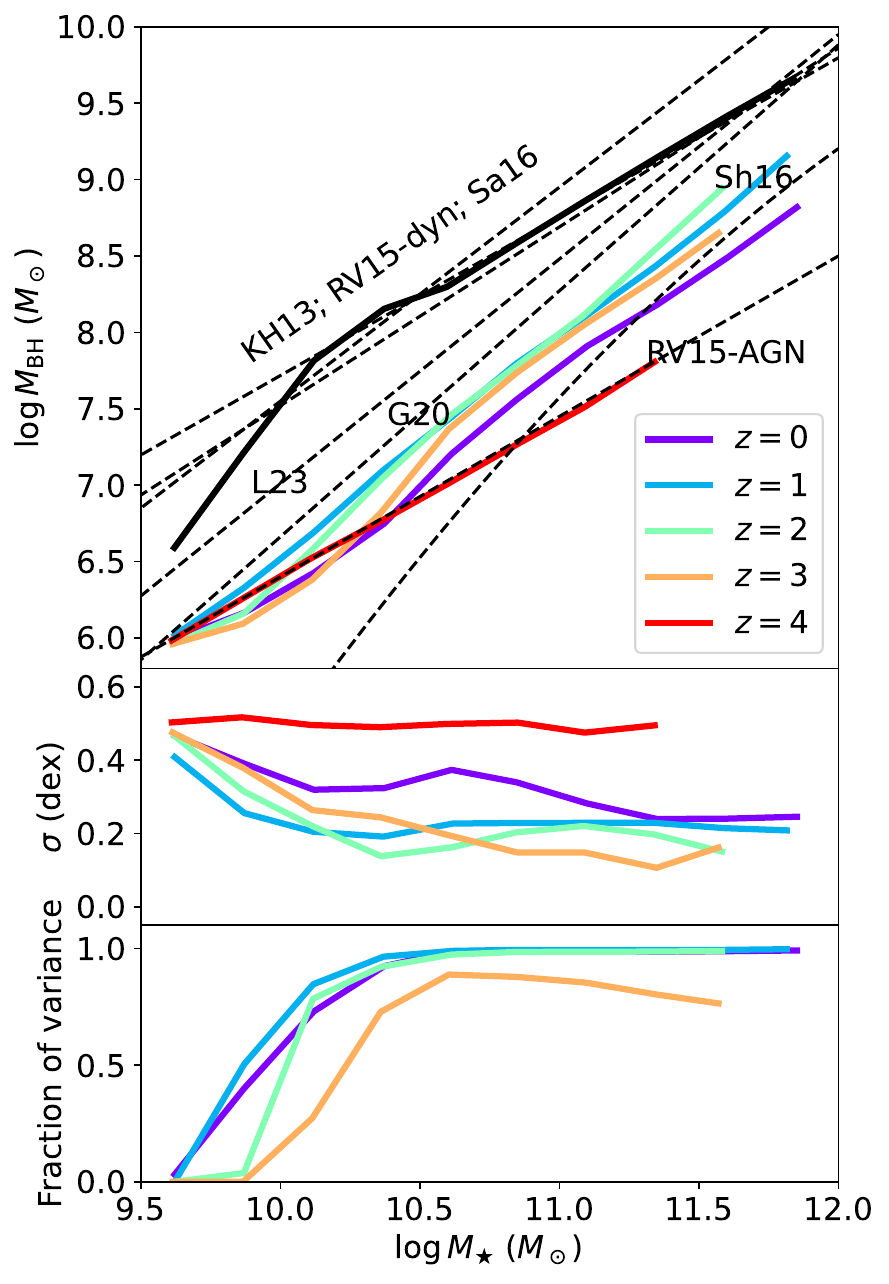}
\caption{Top: Our $M_\mathrm{BH}-M_\star$ relations at different redshifts, plotted as the colored curves. The original TNG-simulated $z=0$ relation is shown as the black solid curve for comparison. We also plot the relations in \citet{Kormendy13}, \citet{Reines15}, \citet{Savorgnan16}, \citet{Shankar16}, \citet{Greene20}, and \citet{Li23} as the black dashed lines, which are labeled as KH13, RV15, Sa16, Sh16, G20, and L23 in the plot, respectively. Two relations in \citet{Reines15} are included -- one based on the SMBH sample with dynamically measured $M_\mathrm{BH}$, and the other based on the AGN sample; these two are labeled as RV15-dyn and RV15-AGN, respectively. The highest three relations are labeled together as ``KH13; RV15-dyn; Sa16'' for a better visualization, and they are all from dynamically measured SMBHs in early-type galaxies. Our recovered relations are apparently lower than the original TNG relation and the early-type dynamically measured ones. Middle: our recovered $1\sigma$ $M_\mathrm{BH}$ scatter in dex versus $M_\star$. Bottom: the fractional contribution of SMBH growth along different $M_\star$ evolutionary tracks to our recovered variance, which is one minus the fraction of variance from seeding scatter.}
\label{fig: scaling}
\end{figure}

Table~\ref{tbl: scaling} presents our predicted $M_\mathrm{BH}-M_\star$ relations at different redshifts based on both TNG100 and TNG300. We plot the TNG300 results in the top panel of Figure~\ref{fig: scaling} and also show the original TNG-simulated $z=0$ relation and the relations in \citet{Kormendy13}, \citet{Reines15}, \citet{Savorgnan16}, \citet{Shankar16}, \citet{Greene20}, and \citet{Li23} for comparison, where the early-type relation is plotted for \citet{Savorgnan16}, and the relation combining both early- and late-type galaxies is adopted for \citet{Greene20}. These literature works were based on different kinds of $M_\mathrm{BH}$ measurements. \citet{Kormendy13}, \citet{Savorgnan16}, \citet{Greene20}, and the higher relation in \citet{Reines15} are derived from dynamical measurements, while \citet{Li23} and the lower relation in \citet{Reines15} are from broad-line AGNs. It is apparent that the dynamically measured ones are generally higher than the AGN-based ones. \citet{Shankar16} argued that the main reason is that the SMBH sample with dynamical $M_\mathrm{BH}$ measurements is biased toward more massive SMBHs, and the higher scaling relations are convolution results of the \textit{debiased} scaling relation, which is also plotted in Figure~\ref{fig: scaling} and closer to the AGN-based relations, with selection biases. It is also worth noting that the highest three relations \citep{Kormendy13, Reines15, Savorgnan16} in Figure~\ref{fig: scaling} are all from early-type galaxies. Late-type galaxies tend to have lower $M_\mathrm{BH}$ at a given $M_\star$ (e.g., \citealt{Savorgnan16, Greene20}). We plot the dynamically measured scaling relation combining both early- and late-type galaxies from \citet{Greene20} in Figure~\ref{fig: scaling}, and its normalization is $>0.5$~dex lower than for early-type galaxies, which helps mitigate the difference between dynamically measured scaling relations and AGN-based ones.\par

\begin{table*}
\caption{The $M_\mathrm{BH}-M_\star$ Scaling Relation}
\label{tbl: scaling}
\centering
\begin{threeparttable}
\begin{tabular}{c|ccccccccc}
\hline
\hline
$\log M_\star$ & $z=0$ & $z=0.5$ & $z=1$ & $z=1.5$ & $z=2$ & $z=2.5$ & $z=3$ & $z=3.5$ & $z=4$\\
\hline
$9.50-9.75$ & 5.99 & 6.00 & 6.02 & 6.00 & 5.96 & 5.96 & 5.99 & 5.95 & 6.00\\
& 5.99 & 5.99 & 6.02 & 5.99 & 5.96 & 5.95 & 5.96 & 5.96 & 5.98\\
$9.75-10.00$ & 6.17 & 6.28 & 6.34 & 6.29 & 6.17 & 6.13 & 6.08 & 6.12 & 6.24\\
& 6.17 & 6.26 & 6.33 & 6.29 & 6.16 & 6.11 & 6.09 & 6.11 & 6.25\\
$10.00-10.25$ & 6.44 & 6.60 & 6.71 & 6.76 & 6.62 & 6.46 & 6.41 & 6.44 & 6.54\\
& 6.43 & 6.55 & 6.69 & 6.79 & 6.58 & 6.43 & 6.38 & 6.42 & 6.52\\
$10.25-10.50$ & 6.84 & 6.96 & 7.14 & 7.21 & 7.06 & 6.90 & 6.84 & 6.89 & 6.80\\
& 6.75 & 6.88 & 7.10 & 7.18 & 7.04 & 6.85 & 6.82 & 6.84 & 6.76\\
$10.50-10.75$ & 7.22 & 7.34 & 7.46 & 7.54 & 7.48 & 7.35 & 7.35 & 7.26 & 7.14\\
& 7.21 & 7.32 & 7.45 & 7.55 & 7.46 & 7.34 & 7.37 & 7.27 & 7.01\\
$10.75-11.00$ & 7.52 & 7.63 & 7.75 & 7.80 & 7.79 & 7.75 & 7.78 & 7.62 & ---\\
& 7.56 & 7.65 & 7.79 & 7.82 & 7.78 & 7.71 & 7.74 & 7.57 & 7.27\\
$11.00-11.25$ & 7.86 & 7.96 & 8.07 & 8.05 & 8.19 & 8.12 & 8.04 & --- & ---\\
& 7.91 & 7.99 & 8.10 & 8.16 & 8.11 & 8.09 & 8.05 & 7.88 & 7.52\\
$11.25-11.50$ & 8.10 & 8.23 & 8.29 & 8.41 & --- & --- & --- & --- & ---\\
& 8.18 & 8.29 & 8.43 & 8.54 & 8.54 & 8.43 & 8.36 & 8.15 & 7.81\\
$11.50-11.75$ & 8.36 & 8.55 & --- & --- & --- & --- & --- & --- & ---\\
& 8.49 & 8.59 & 8.79 & 8.93 & 8.94 & 8.83 & 8.65 & --- & ---\\
$11.75-12.00$ & --- & --- & --- & --- & --- & --- & --- & --- & ---\\
& 8.82 & 9.00 & 9.16 & 9.37 & --- & --- & --- & --- & ---\\
\hline
\hline
\end{tabular}
\begin{tablenotes}
\item
\textit{Notes.} The first column lists the $\log M_\star$ range of each bin, and the other columns show the median $\log M_\mathrm{BH}$ values at the corresponding bins. Values in bins with less than 20 sources are not shown. Within each cell, the top and bottom parts show the TNG100 and TNG300 results, respectively.
\end{tablenotes}
\end{threeparttable}
\end{table*}

The figure shows that our $z=0$ scaling relation is more similar to the AGN-based ones. This is not due to the fact that our initial SMBH seeding follows the AGN-based relation in \citet{Reines15}. We have tried adopting the higher dynamically measured relation as the seeding relation, and the recovered scaling relation is still below the dynamically measured one, as will be shown in Section~\ref{sec: differentseed}. The main reason is that, even if we begin with the dynamically measured relation, the subsequent accretion is not sufficient to maintain the high normalization. Therefore, our \mbox{X-ray}-based $\dot{M}_a$ from \citet{Zou24} is inconsistent with the dynamically measured relation. The same finding was also shown in \citet{Shankar20a}. This may be because \mbox{X-ray} surveys miss some accretion power or the scaling relation is overestimated. However, to match the scaling relation by increasing $\dot{M}_a$, we need to elevate it by one order of magnitude. We do not think this level of uncertainty can be from bolometric corrections or radiative efficiency (Section~\ref{sec: accretion}), and it is hard to explain how the majority of the accretion power could be missed by \mbox{X-rays}. On the other hand, a more plausible reason, as proposed in \citet{Shankar16, Shankar20a, Shankar20b}, is that the dynamically measured SMBH sample is biased toward more massive SMBHs. Nevertheless, it should be noted that this claim is still under debate, and some works also argued that the AGN-based relation in \citet{Reines15} or even the higher dynamically measured one may underestimate the number of SMBHs at the massive end (e.g., \citealt{Pesce21, SatoPolito23}).\par
\begin{figure}
\includegraphics[width=\hsize]{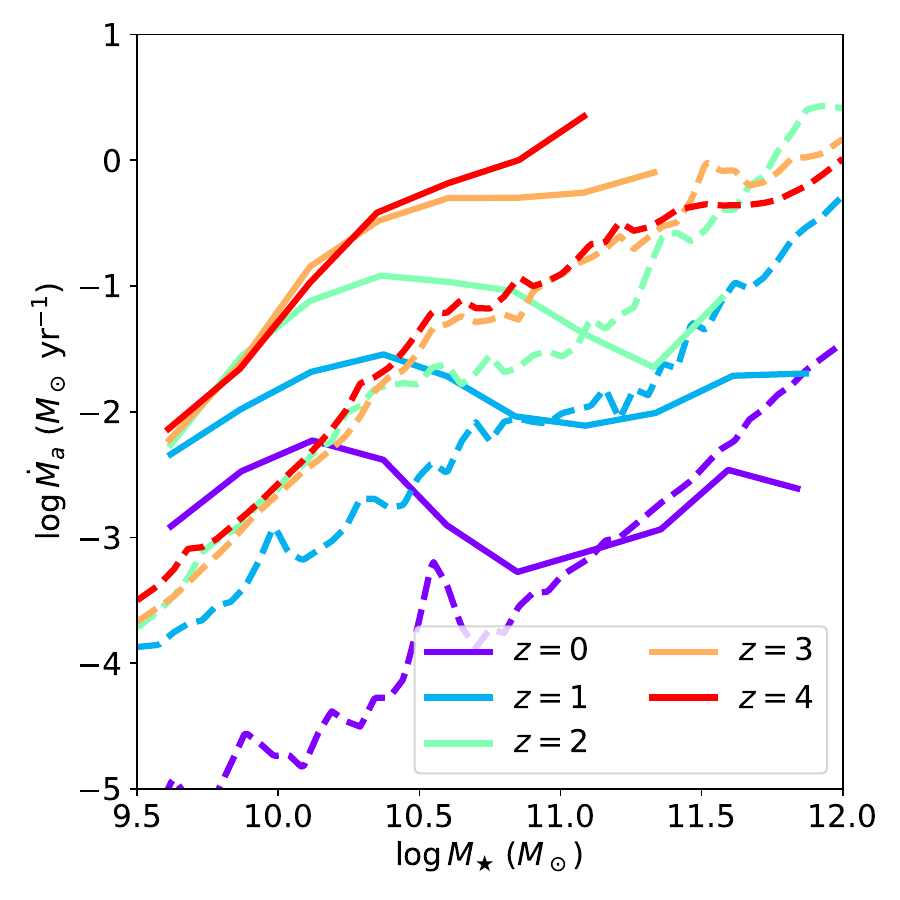}
\caption{$\dot{M}_a$ versus $M_\star$ at different redshifts. The solid and dashed curves are from the original TNG300 simulation and the observed relation in \citet{Zou24}, respectively. There are orders-of-magnitude differences between the simulated and observed $\dot{M}_a$, and the simulated $\dot{M}_a$ often lacks the positive correlation with $M_\star$.}
\label{fig: bhar_tng}
\end{figure}

Besides, the original TNG-simulated relation is consistent with the higher dynamically measured relations, which is expected because TNG is tuned to match these relations. The discrepancy between our post-processed results and the original TNG ones originates from the fact that the simulated $\dot{M}_a$ is greatly different from observations. We compare the original simulated $\dot{M}_a$, derived by averaging the \texttt{SubhaloBHMdot} column in the TNG catalogs in each $M_\star$ bin, and the observed $\dot{M}_a$ in \citet{Zou24} as a function of $M_\star$ at different redshifts in Figure~\ref{fig: bhar_tng}. The figure reveals that TNG can overestimate $\dot{M}_a$ by orders of magnitude, especially at $M_\star\lesssim10^{11}~M_\odot$. Besides, the simulated $\dot{M}_a$ does not further increase with $M_\star$ at $M_\star\gtrsim10^{10.5}~M_\odot$ and $z\lesssim2$ due to the suppression of accretion \citep{Weinberger18}, which contradicts observations. Therefore, even if the original TNG simulations can recover a plausible $M_\mathrm{BH}-M_\star$ scaling relation, which is an integrated metric, its simulated accretion channel is still generally unrealistic \citep{Habouzit21, Habouzit22}. This is not surprising because TNG was not calibrated with the AGN population. Given this, our post-processing of the TNG data with direct observational constraints added can improve the SMBH population compared to the original purely simulation-based results. Also, the comparison here is mainly qualitative, and more quantitative comparisons require constructing detailed mock observations from the TNG simulations to convert the TNG results into the space of direct observables (e.g., \citealt{Habouzit22}), which is beyond the scope of this work. Besides, note that the limitation of the simulated accretion does not directly propagate into the simulated galaxy properties in the sense that galaxies are instead regulated by the feedback, whose parameters can be tuned independently from the accretion model. Given the fact that the simulated galaxy population has been calibrated with several metrics (e.g., the SMF, the cosmic star formation rate density, the correlation between $M_\star$ and halo mass, the halo gas fraction, and the galaxy size; see \citealt{Pillepich18b}), the simulated galaxies are generally more reliable than the simulated SMBHs.\par

Our scaling relation in Figure~\ref{fig: scaling} does not evolve much from $z=4$ to $z=0$, with a difference $\lesssim0.5$~dex. The $z=0$ curve is slightly systematically below the higher-redshift ones, but the difference is still small ($<0.3$~dex). Such small evolution is generally supported by recent works, including both observational studies and cosmological simulations (e.g., \citealt{Yang18, Yang19, Yang22, Shankar20a, Suh20, Sun15, Habouzit21, Li21a, Li21b, Li23}).\par
Some parts of the scaling relations are dominated by our initial SMBH seeding; for example, our $z=4$ curve, by design, exactly follows \citet{Reines15} with perturbations from the seeding scatter. The contributions of our adopted seed mass to $M_\mathrm{BH}$ will be explored in greater detail in Section~\ref{sec: massassembly}. Briefly, SMBHs in massive galaxies with $M_\star\gtrsim10^{10}~M_\odot$ at $z=3$ have already lost their memory about their seed masses at $z=4$, while SMBHs in low-mass galaxies with $M_\star\lesssim10^{10}~M_\odot$ at $z=0$ still retain significant information at high redshifts. Therefore, the low-mass parts of our scaling relations in Figure~\ref{fig: scaling} are mainly set by our initial conditions at all redshifts, while the more massive parts ($M_\star\gtrsim10^{10}~M_\odot$) at $z\leq3$ are instead genuinely created by the SMBH growth. Therefore, our results in the massive parts should be robust as long as the accumulated SMBH masses before $z=4$ are negligible. We will present the SMBH growth rates via different channels in Section~\ref{sec: channelgrow}, where we will show that the SMBH growth is mainly dominated by the accretion channel in most cases, and thus our scaling relations above $M_\star\gtrsim10^{10}~M_\odot$ are driven by the correlation between $\dot{M}_a$ and $(M_\star, z)$. This is consistent with \citet{Yang18} and \citet{Shankar20a}, who neglected the merger channel, integrated $\dot{M}_a(M_\star, z)$, and found that the local-universe scaling relation can apparently be recovered and is close to the lower, AGN-based scaling relation.\par
It is also worth noting that, if we adopt a higher seeding relation such as the dynamically measured ones, the recovered scaling relation would increase with increasing redshift, as will be shown in more detail in Section~\ref{sec: differentseed}. This is because the high seeding scaling-relation normalization cannot be maintained by the subsequent accretion, and thus the overall normalization gradually decreases as the Universe ages. However, such a redshift-dependent evolution requires SMBHs to accumulate at least half of their masses by $z=4$ (see Section~\ref{sec: differentseed}), which contradicts previous findings that the accumulation of half of the $M_\mathrm{BH}$ occurred much later; see \citet{Vito16, Vito18}, \citet{Shankar20a}, \citet{Shen20}, \citet{Habouzit21}, and \citet{Zhang23} for examples from a variety of observational or theoretical perspectives; but also note that recent JWST results tend to favor stronger early SMBH growth with over-massive SMBHs at $z>4$ (e.g., \citealt{Maiolino24, Pacucci23}). This redshift dependence of the scaling relation also contradicts observations that no apparent evolution is seen with respect to redshift (e.g., Figure~10 in \citealt{Li23}).\par
In the middle panel of Figure~\ref{fig: scaling}, we plot our recovered $1\sigma$ scatter of the scaling relation versus $M_\star$. Our recovered $M_\mathrm{BH}$ scatter is contributed by two components -- one is from the seeding scatter, and the other is from the SMBH growth along diverse $M_\star$ evolutionary tracks. To assess the relative contributions from these two components, we rerun our analyses in Section~\ref{sec: methodology} but set $\dot{M}_a=0$, and the resulting $M_\mathrm{BH}$ only keeps the seeding information, as will be elaborated more in Section~\ref{sec: massassembly}. We run it twice so that we can calculate the contribution of the seeding scatter to the total $M_\mathrm{BH}$ variance based on the dispersion of the difference in $M_\mathrm{BH}(\dot{M}_a=0)$ between these two independent runs.\footnote{Mathematically, $M_\mathrm{BH}(\dot{M}_a=0)=\overline{M_\mathrm{BH}}(\dot{M}_a=0)+\varepsilon_\mathrm{seed}$, where the first term is the mean $M_\mathrm{BH}(\dot{M}_a=0)$, and the second term is a random variable with a mean of 0 originating from the seeding scatter. We do not directly assess $\overline{M_\mathrm{BH}}(\dot{M}_a=0)$ because it cannot be measured with a simple run without seeding scatter. The reason is that the seeding scatter is added in the logarithmic space, and eliminating the scatter would change $\overline{M_\mathrm{BH}}(\dot{M}_a=0)$ due to the fact that the mean of a lognormal distribution centered at 0 is not 0. Instead, we use two independent runs with seeding scatters to subtract this term. For these two realizations, we have $\delta M_\mathrm{BH}=M_\mathrm{BH}(\mathrm{realization~1})-M_\mathrm{BH}(\mathrm{realization~2})=\varepsilon_\mathrm{seed}(\mathrm{realization~1})-\varepsilon_\mathrm{seed}(\mathrm{realization~2})$. Thus, $\mathrm{Var}(\delta M_\mathrm{BH})=2\mathrm{Var}(\varepsilon_\mathrm{seed})$. Each SMBH has a $\delta M_\mathrm{BH}$ value that can be regarded as one realization of the underlying population $\delta M_\mathrm{BH}$ distribution. Thus, the typical contribution from the seeding scatter to $M_\mathrm{BH}$, $\mathrm{Var}(\varepsilon_\mathrm{seed})$, can be measured from the dispersion of $\delta M_\mathrm{BH}$.} We then present the fractional contribution of variance from the second component, the SMBH growth along diverse $M_\star$ evolutionary tracks, in the bottom panel of Figure~\ref{fig: scaling}. The figure shows that our recovered variance is dominated by this second component at $M_\star>10^{10}~M_\odot$ and $z<3$, indicating the memory retained about $M_\mathrm{BH}$ seed masses quickly fades away as $M_\star$ increases from $10^{9.5}~M_\odot$ to $10^{10}~M_\odot$. We focus on the low-memory regime here, where our predicted local-universe scatter is $\approx0.3-0.4$~dex. The observationally measured intrinsic scatter is $\approx0.5$~dex \citep{Reines15, Li23}. It should be noted that our scatters are only lower limits because we assumed that all the SMBHs exactly follow the $\dot{M}_a(M_\star, z)$ function; however, there is an additional variance in $\dot{M}_a$ not captured by the $(M_\star, z)$ predictor set, such as the residual dependence on star formation rate and morphology (e.g., compactness). Nevertheless, the ratio between our variance and the observed one reaches $\approx50\%$, indicating that the SMBH growth along diverse $M_\star$ evolutionary tracks can already explain half of the observed $M_\mathrm{BH}-M_\star$ variance. The middle panel of Figure~\ref{fig: scaling} also indicates that the scatter may decrease from $z=0$ to $z=1$ and remains roughly constant up to $z=3$, but this conclusion is subject to the amount of variance not covered by our $\dot{M}_a(M_\star, z)$ function.

% BHMF
\subsection{BHMF}
\label{sec: bhmf}
The BHMF, denoted as $\phi(M_\mathrm{BH})$, is the number of SMBHs per unit comoving volume per unit $\log M_\mathrm{BH}$. Although this quantity is fundamental, it is a nontrivial challenge to measure it reliably because indirect tracers of $M_\mathrm{BH}$ are needed; see, for example, \citet{Kelly12} for a review. With our constructed SMBH population, we can compute the BHMF at different redshifts. We do not include wandering SMBHs in our BHMF because usually only central SMBHs are observable, and the BHMF in previous literature also refers to central SMBHs. We will focus on these wandering SMBHs in Section~\ref{sec: wander}. Given our $M_\star$ cut at $10^{9.5}~M_\odot$, which corresponds to $M_\mathrm{BH}=7\times10^5~M_\odot$ according to the scaling relation in \citet{Reines15}, we focus on the BHMF with $M_\mathrm{BH}>10^6~M_\odot$. It is possible that some dwarf galaxies contain SMBHs above this threshold, and thus our BHMF may be slightly underestimated at low $M_\mathrm{BH}$, but this bias should be negligible when $M_\mathrm{BH}\gtrsim10^7~M_\odot$.\par
\begin{figure}
\includegraphics[width=\hsize]{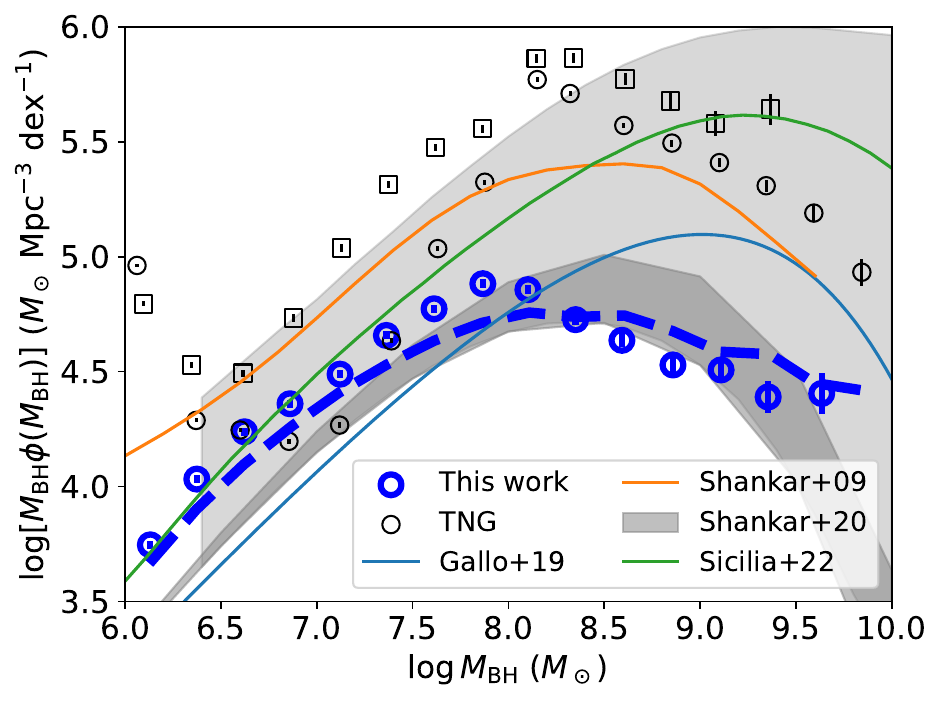}
\caption{The BHMF at $z=0$. We show $M_\mathrm{BH}\phi(M_\mathrm{BH})$ instead of $\phi(M_\mathrm{BH})$ to illustrate the typical $M_\mathrm{BH}$ range where most SMBH mass resides. The blue open circles with $1\sigma$ error bars are our BHMFs, where the uncertainties only represent the limited number of sources included in the simulation boxes. The blue dashed line is the resulting BHMF with an additional scatter of 0.5~dex added to our $M_\mathrm{BH}$. The black open circles and squares with $1\sigma$ error bars are the BHMFs in the original TNG300 and TNG100 simulations. The BHMFs in \citet{Gallo19}, \citet{Shankar09, Shankar20b}, and \citet{Sicilia22} are also plotted, as labeled in the legend. The lighter shaded region shows the uncertainty range in \citet{Shankar20b} considering the uncertainties of scaling relations between $M_\mathrm{BH}$ and galaxy properties (see also \citealt{Sicilia22}), and the darker shaded region is the BHMF in \citet{Shankar20b} based on the scaling relation after correcting for the selection bias of dynamically measured SMBHs.}
\label{fig: bhmf_z0}
\end{figure}

We bin our sources with a $M_\mathrm{BH}$ step of 0.25~dex to measure the BHMF, and only those bins with at least 20 sources are used. The BHMF at different redshifts is tabulated in Table~\ref{tbl: bhmf} for both TNG100 and TNG300. Figure~\ref{fig: bhmf_z0} presents $M_\mathrm{BH}\phi(M_\mathrm{BH})$ at $z=0$. We show $M_\mathrm{BH}\phi(M_\mathrm{BH})$, the cosmic SMBH mass density per unit $\log M_\mathrm{BH}$, instead of $\phi(M_\mathrm{BH})$ to illustrate the typical $M_\mathrm{BH}$ range where most SMBH mass resides. The massive end of the BHMF may visually look like a plateau, which is primarily due to statistical fluctuations. If we further plot the BHMF to higher masses, the BHMF would continue to decrease. However, it is worth noting that our massive end generally does not decay as fast as other works. This originates from the fact that the TNG SMF also does not decay sufficiently fast at the very massive end \citep{Pillepich18a}. Our uncertainties are calculated as the Poisson error of the number of sources in each bin following \citet{Gehrels86} and do not include other systematic uncertainties, which could be larger. There are also additional uncertainties not accounted for from the statistical uncertainties of the $\dot{M}_a(M_\star, z)$ function, but Appendix~\ref{append: adderr} shows that these terms should be small in general. There are several sources of systematic uncertainties. For example, as mentioned in Section~\ref{sec: accretion}, there is a typical systematic uncertainty up to a factor of a few when measuring $\dot{M}_a$ from \mbox{X-ray} emission. Another source of systematic uncertainties is from TNG, i.e., how real the simulated universe is. Different models can produce different evolutionary tracks of $M_\star$ for galaxies (e.g., see Figure~3 in \citealt{Shankar20a}). Besides, $\dot{M}_a(M_\star, z)$ only captures part of the variance in SMBH growth, and our recovered scatter of the $M_\mathrm{BH}-M_\star$ scaling relation in Section~\ref{sec: scaling} is smaller than the observed one. We also tried adding an additional scatter of 0.5~dex to $M_\mathrm{BH}$ and systematically reducing all $M_\mathrm{BH}$ by 0.29~dex to keep the total mass unchanged.\footnote{To ensure the mean of a lognormal distribution is 0, its center should be negative, as reflected by our leftward shift of $M_\mathrm{BH}$.} This additional scatter increases the total scatter of $M_\mathrm{BH}$ at a given $M_\star$ to be $0.6-0.7$~dex. The new BHMF is plotted as the blue dashed curve in Figure~\ref{fig: bhmf_z0}, which presents a slightly heavier massive tail.\par

\begin{table*}
\caption{The BHMF}
\label{tbl: bhmf}
\centering
\begin{threeparttable}
\begin{tabular}{c|ccccccccc}
\hline
\hline
$\log M_\mathrm{BH}$ & $z=0$ & $z=0.5$ & $z=1$ & $z=1.5$ & $z=2$ & $z=2.5$ & $z=3$ & $z=3.5$ & $z=4$\\
\hline
$6.00-6.25$ & $-2.28$ & $-2.33$ & $-2.41$ & $-2.50$ & $-2.58$ & $-2.68$ & $-2.81$ & $-3.01$ & $-3.30$\\
& $-2.39$ & $-2.42$ & $-2.50$ & $-2.58$ & $-2.62$ & $-2.72$ & $-2.86$ & $-3.07$ & $-3.31$\\
$6.25-6.50$ & $-2.26$ & $-2.30$ & $-2.39$ & $-2.45$ & $-2.57$ & $-2.65$ & $-2.85$ & $-3.10$ & $-3.31$\\
& $-2.34$ & $-2.36$ & $-2.44$ & $-2.54$ & $-2.63$ & $-2.70$ & $-2.88$ & $-3.09$ & $-3.32$\\
$6.50-6.75$ & $-2.33$ & $-2.30$ & $-2.39$ & $-2.54$ & $-2.65$ & $-2.76$ & $-2.98$ & $-3.17$ & $-3.40$\\
& $-2.38$ & $-2.37$ & $-2.45$ & $-2.61$ & $-2.69$ & $-2.80$ & $-2.99$ & $-3.21$ & $-3.37$\\
$6.75-7.00$ & $-2.44$ & $-2.45$ & $-2.49$ & $-2.58$ & $-2.72$ & $-2.93$ & $-3.11$ & $-3.35$ & $-3.43$\\
& $-2.50$ & $-2.51$ & $-2.54$ & $-2.60$ & $-2.75$ & $-2.96$ & $-3.17$ & $-3.34$ & $-3.51$\\
$7.00-7.25$ & $-2.59$ & $-2.57$ & $-2.56$ & $-2.67$ & $-2.84$ & $-3.10$ & $-3.33$ & $-3.48$ & $-3.71$\\
& $-2.63$ & $-2.60$ & $-2.57$ & $-2.69$ & $-2.87$ & $-3.12$ & $-3.36$ & $-3.52$ & $-3.70$\\
$7.25-7.50$ & $-2.69$ & $-2.65$ & $-2.62$ & $-2.71$ & $-2.94$ & $-3.24$ & $-3.56$ & $-3.80$ & $-4.07$\\
& $-2.71$ & $-2.68$ & $-2.68$ & $-2.75$ & $-2.96$ & $-3.25$ & $-3.47$ & $-3.64$ & $-3.91$\\
$7.50-7.75$ & $-2.84$ & $-2.79$ & $-2.81$ & $-2.84$ & $-3.08$ & $-3.46$ & $-3.73$ & $-3.95$ & ---\\
& $-2.84$ & $-2.80$ & $-2.83$ & $-2.88$ & $-3.09$ & $-3.37$ & $-3.61$ & $-3.86$ & $-4.19$\\
$7.75-8.00$ & $-3.03$ & $-2.96$ & $-2.99$ & $-3.16$ & $-3.45$ & $-3.78$ & $-3.92$ & --- & ---\\
& $-2.98$ & $-2.93$ & $-2.94$ & $-3.06$ & $-3.29$ & $-3.60$ & $-3.79$ & $-4.20$ & $-4.52$\\
$8.00-8.25$ & $-3.33$ & $-3.33$ & $-3.39$ & $-3.57$ & $-3.78$ & $-3.97$ & --- & --- & ---\\
& $-3.25$ & $-3.22$ & $-3.27$ & $-3.42$ & $-3.64$ & $-3.86$ & $-4.17$ & $-4.49$ & $-4.94$\\
$8.25-8.50$ & $-3.69$ & $-3.73$ & $-3.76$ & $-3.85$ & $-4.09$ & --- & --- & --- & ---\\
& $-3.62$ & $-3.61$ & $-3.66$ & $-3.79$ & $-4.04$ & $-4.24$ & $-4.40$ & $-4.94$ & $-5.23$\\
$8.50-8.75$ & $-4.07$ & $-4.07$ & $-4.15$ & --- & --- & --- & --- & --- & ---\\
& $-3.96$ & $-4.00$ & $-4.08$ & $-4.25$ & $-4.38$ & $-4.66$ & $-4.88$ & --- & ---\\
$8.75-9.00$ & --- & --- & --- & --- & --- & --- & --- & --- & ---\\
& $-4.33$ & $-4.34$ & $-4.40$ & $-4.56$ & $-4.77$ & $-4.98$ & --- & --- & ---\\
$9.00-9.25$ & --- & --- & --- & --- & --- & --- & --- & --- & ---\\
& $-4.60$ & $-4.67$ & $-4.79$ & $-4.97$ & $-5.06$ & --- & --- & --- & ---\\
$9.25-9.50$ & --- & --- & --- & --- & --- & --- & --- & --- & ---\\
& $-4.97$ & $-5.02$ & $-5.19$ & $-5.40$ & --- & --- & --- & --- & ---\\
$9.50-9.75$ & --- & --- & --- & --- & --- & --- & --- & --- & ---\\
& $-5.23$ & $-5.32$ & $-5.42$ & --- & --- & --- & --- & --- & ---\\
\hline
\hline
\end{tabular}
\begin{tablenotes}
\item
\textit{Notes.} The first column lists the $\log M_\mathrm{BH}$ range of each bin, and the other columns show $\log\phi(M_\mathrm{BH})$ in $\mathrm{Mpc^{-3}~dex^{-1}}$ in the corresponding bins. Values in bins with less than 20 sources are not shown. Within each cell, the top and bottom parts show the TNG100 and TNG300 results, respectively.
\end{tablenotes}
\end{threeparttable}
\end{table*}

We also show the BHMFs in \citet{Shankar09, Shankar20b}, \citet{Gallo19}, \citet{Sicilia22}, and the original TNG100 and TNG300 simulations as comparisons. The former three are all derived from scaling relations of $M_\mathrm{BH}$ with galaxy properties, while the latter three are instead theoretical ones. The lighter shaded region in Figure~\ref{fig: bhmf_z0} is from \citet[see also \citealt{Sicilia22}]{Shankar20b} and reflects the overall uncertainty of the scaling relation and also subdominant Poisson uncertainties of limited source counts. Its upper bound is based on the dynamically measured scaling relation, while its lower bound is derived by correcting the selection bias that more massive SMBHs are more likely to have dynamically measured $M_\mathrm{BH}$, which, hence, contains an assumption that the observed scaling relation is highly biased, as discussed in Section~\ref{sec: scaling}. We also explicitly plot the debiased BHMF in \citet{Shankar20b} as the darker shaded region in Figure~\ref{fig: bhmf_z0}. The typical BHMF uncertainty range originating from the scaling relation (i.e., biased or not) is $\approx1$~dex below $10^8~M_\odot$ but increases to several orders of magnitude at the massive end where the BHMF exponentially decays.\par
Our BHMF is more consistent with the debiased BHMF in \citet{Shankar20b}, echoing our finding in Section~\ref{sec: scaling} that our recovered scaling relation is lower than the dynamically measured relation. However, as Section~\ref{sec: differentseed} will show, the BHMF could be elevated, especially at the massive end, if we seed SMBHs with higher $M_\mathrm{BH}$, but this requires the SMBH growth before $z=4$ to have a considerable fractional contribution to the final $M_\mathrm{BH}$ at $z=0$. Our BHMF also shows a notable difference from the original TNG simulated SMBHs, as similarly shown in Section~\ref{sec: scaling}.\par
Integrating the BHMF, we obtain the local SMBH mass density to be $\rho_\mathrm{BH}=1.3\times10^5~M_\odot~\mathrm{Mpc}^{-3}$ and $1.5\times10^5~M_\odot~\mathrm{Mpc}^{-3}$ for TNG100 and TNG300, respectively. These numbers are in agreement with recent findings in \citet{Shankar20b}, who derived $\rho_\mathrm{BH}=(1-2)\times10^5~M_\odot~\mathrm{Mpc}^{-3}$. Earlier results based on the local scaling relations between $M_\mathrm{BH}$ and galaxy properties, which tend to be higher than the debiased relation, may return slightly larger $\rho_\mathrm{BH}=(3-6)\times10^5~M_\odot~\mathrm{Mpc}^{-3}$ (e.g., \citealt{Graham07, Brandt15} and references therein). By assuming a radiative efficiency of 0.1 and integrating the bolometric quasar luminosity function from high redshifts to $z=0$, \citet{Hopkins07} and \citet{Shen20} derived $\rho_\mathrm{BH}=5\times10^5~M_\odot~\mathrm{Mpc}^{-3}$. We also integrated the AGN \mbox{X-ray} luminosity function in \citet{Ueda14}, adopting the bolometric correction from the \mbox{X-ray} luminosity to the bolometric luminosity in \citet{Duras20}, and obtained a lower $\rho_\mathrm{BH}=(2-3)\times10^5~M_\odot~\mathrm{Mpc}^{-3}$. Therefore, there is an uncertainty of a factor of a few in $\rho_\mathrm{BH}$, depending upon the detailed methodology. Our values, similar to \citet{Shankar20b}, are close to the lower bound compared to previous works.\par
\begin{figure}
\includegraphics[width=\hsize]{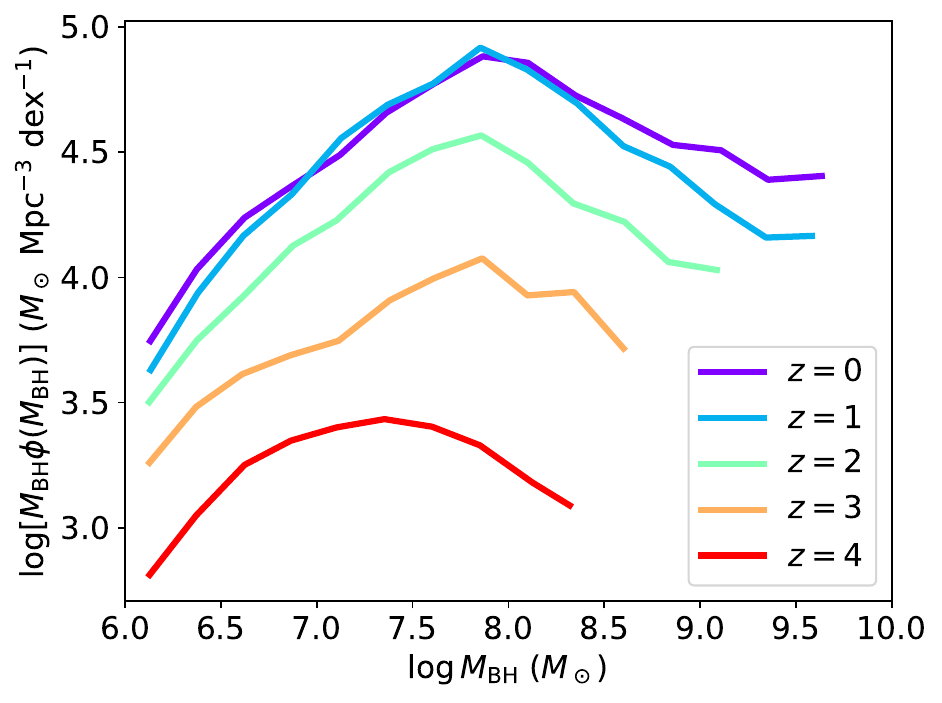}
\caption{$M_\mathrm{BH}\phi(M_\mathrm{BH})$ at different redshifts.}
\label{fig: bhmf_evo}
\end{figure}

We further explore the redshift evolution of the BHMF in Figure~\ref{fig: bhmf_evo}. The BHMF at $z=4$ reflects our initial condition and is nothing more than the corresponding TNG SMF scaled down by the scaling relation in \citet{Reines15}. From $z=4$ to $z=1$, the BHMF normalization increases steadily at all $M_\mathrm{BH}$, with limited changes in the overall shape. From $z=1$ to $z=0$, the BHMF almost remains the same, indicating that the overall BHMF has been largely built by $z=1$. Although BHMFs in previous literature can agree with each other within $\approx1$~dex at $z=0$, there have been large differences in how the BHMF is built over time. \citet{Shankar09} concluded that the BHMF across the whole $M_\mathrm{BH}$ range evolved strongly between $z=4$ and $z=1$, and its massive part at $M_\mathrm{BH}\gtrsim10^8~M_\odot$ was mostly built up by $z=1$, but there would still be large evolution by $\approx1$~dex for the less-massive part from $z=1$ to $z=0$. \citet{Merloni08} predicted that only the massive part of the BHMF at $M_\mathrm{BH}\gtrsim10^{7.5}~M_\odot$ evolved before $z=1$, while the less-massive part began its evolution at $z\lesssim1$ and gradually caught up to the BHMF at $z=0$. Recent work in \citet{Sicilia22} predicts an evolution pattern more similar to ours, where the normalization of the BHMF steadily increases from $z=4$ to $z=1$ and then mostly locks down with little evolution down to $z=0$. Cosmological simulations also predict strong BHMF evolution from $z=4$ to $z=1$ at any $M_\mathrm{BH}$ and further accumulation of massive SMBHs with $M_\mathrm{BH}\gtrsim10^9~M_\odot$ after $z=1$, but there are divergences regarding whether the BHMF below $10^9~M_\odot$ continues to significantly evolve at $z<1$ \citep{Habouzit21}.\par

\begin{figure}
\includegraphics[width=\hsize]{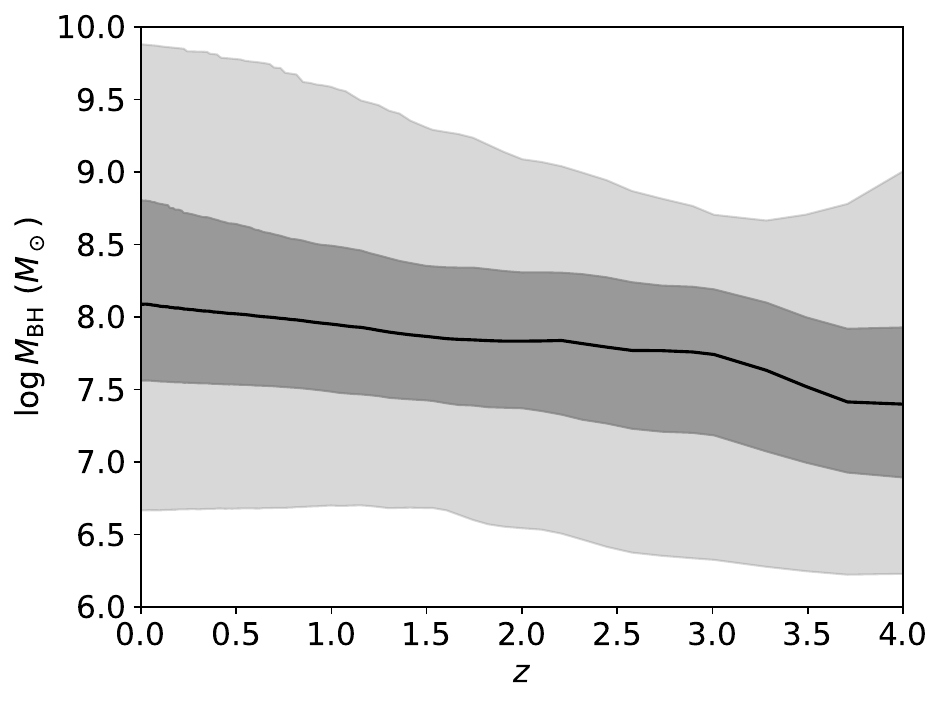}
\caption{The $M_\mathrm{BH}$ values below which 5\%, 25\%, 50\%, 75\%, and 95\% of all the SMBH mass is included as functions of $z$. The black curve represents the 50\% one, and the shaded region encloses 25\%-75\% and 5\%-95\% mass ranges.}
\label{fig: massquantiles}
\end{figure}

We further show the $M_\mathrm{BH}$ values below which 5\%, 25\%, 50\%, 75\%, and 95\% of all the SMBH mass is included as functions of $z$ in Table~\ref{tbl: massquantiles} and Figure~\ref{fig: massquantiles}. The typical $M_\mathrm{BH}\approx10^8~M_\odot$ at $z=0$, below (above) which half of all the SMBH mass is distributed. These numbers only decrease modestly with increasing redshift because the overall BHMF shape remains similar at different redshifts (Figure~\ref{fig: bhmf_evo}).

\begin{table}
\caption{Typical $M_\mathrm{BH}$ at Different Redshifts}
\label{tbl: massquantiles}
\centering
\begin{threeparttable}
\begin{tabular}{c|ccccc}
\hline
\hline
$z$ & 5\% & 25\% & 50\% & 75\% & 95\%\\
\hline
0.0 & 6.52 & 7.37 & 7.88 & 8.44 & 9.28\\
& 6.67 & 7.56 & 8.09 & 8.81 & 9.88\\
0.5 & 6.55 & 7.36 & 7.83 & 8.29 & 9.11\\
& 6.68 & 7.53 & 8.02 & 8.64 & 9.78\\
1.0 & 6.57 & 7.34 & 7.77 & 8.22 & 8.97\\
& 6.70 & 7.48 & 7.95 & 8.49 & 9.59\\
1.5 & 6.53 & 7.29 & 7.69 & 8.09 & 8.78\\
& 6.68 & 7.42 & 7.87 & 8.35 & 9.31\\
2.0 & 6.42 & 7.18 & 7.61 & 8.09 & 8.68\\
& 6.54 & 7.37 & 7.83 & 8.31 & 9.09\\
2.5 & 6.28 & 7.01 & 7.56 & 8.04 & 8.51\\
& 6.40 & 7.25 & 7.78 & 8.26 & 8.91\\
3.0 & 6.22 & 6.93 & 7.53 & 8.00 & 8.35\\
& 6.33 & 7.18 & 7.74 & 8.19 & 8.71\\
3.5 & 6.15 & 6.84 & 7.35 & 7.79 & 8.28\\
& 6.25 & 6.99 & 7.52 & 7.99 & 8.71\\
4.0 & 6.17 & 6.81 & 7.21 & 7.74 & 8.55\\
& 6.23 & 6.89 & 7.40 & 7.93 & 9.00\\
\hline
\hline
\end{tabular}
\begin{tablenotes}
\item
\textit{Notes.} The $M_\mathrm{BH}$ values below which 5\%, 25\%, 50\%, 75\%, and 95\% of all the SMBH mass is included at different redshifts. The first column is the redshift, and the remaining ones list the corresponding $\log M_\mathrm{BH}$. Within each cell, the top and bottom parts show the TNG100 and TNG300 results, respectively.
\end{tablenotes}
\end{threeparttable}
\end{table}

% Growth rate
\subsection{Growth Rates in Different Channels}
\label{sec: channelgrow}
The accretion-driven $M_\mathrm{BH}$ growth rate, $\dot{M}_a$, is from \citet{Zou24}, and we further measure $\dot{M}_m$ and compare it with $\dot{M}_a$ in this section. For a given SMBH, we consider all of its immediate progenitors in its merger tree and label the most massive one as the primary one, and the mass added to this primary one is from all the remaining SMBHs. This added mass is then divided by the corresponding time interval to calculate $\dot{M}_m$. Note that the actual added mass depends on $p_\mathrm{merge}$ and the GW-radiation mass loss, as discussed in Section~\ref{sec: merger}.\par
\begin{figure}
\includegraphics[width=\hsize]{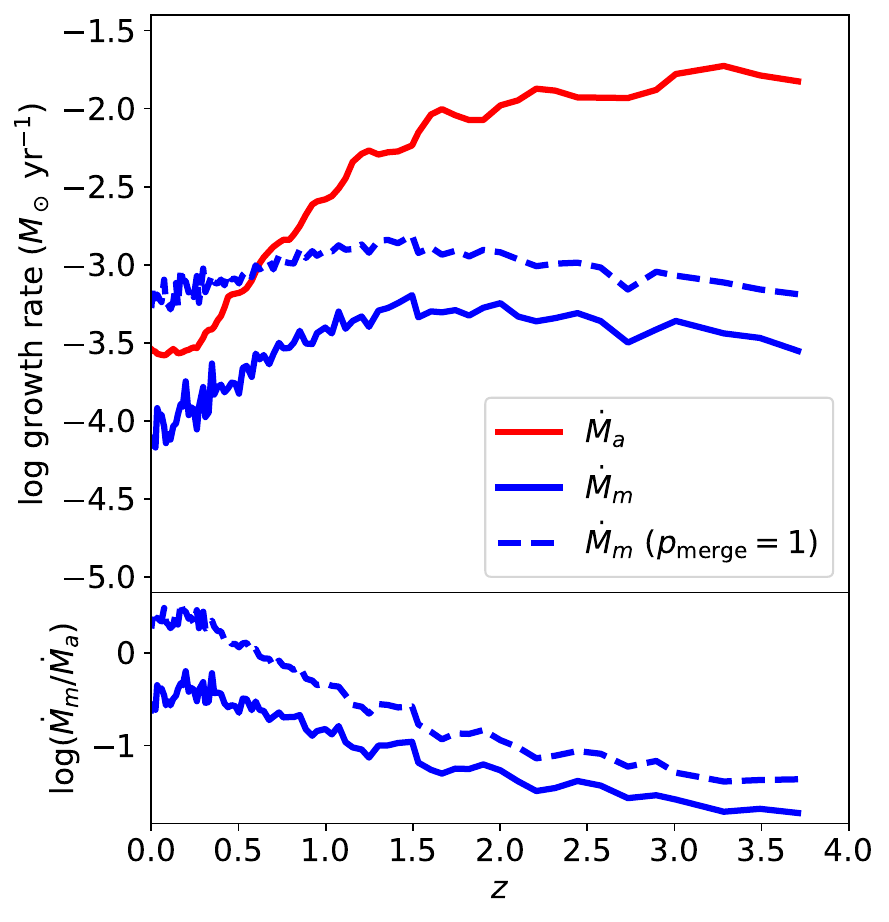}
\caption{Top: The mean growth rates in the accretion channel (red) and the merger channel (blue) for our whole sample. The solid and dashed merger-channel curves correspond to the $p_\mathrm{merge}$ in \citet{Tremmel18} and $p_\mathrm{merge}=1$, respectively, and the latter represents the upper limit of $\dot{M}_m$. Bottom: $\log(\dot{M}_m/\dot{M}_a)$ versus $z$. The visually stronger fluctuations at low redshifts are due to the decreasing redshift interval length between snapshots with the same time interval.}
\label{fig: gr_mean}
\end{figure}

We compare $\dot{M}_a$ and $\dot{M}_m$ averaged over all of our TNG sources in Figure~\ref{fig: gr_mean}. Two types of $\dot{M}_m$ with different $p_\mathrm{merge}$ are included; one is for our previously adopted $p_\mathrm{merge}$ in \citet{Tremmel18}, and the other is $p_\mathrm{merge}=1$, which represents the upper limit for $\dot{M}_m$. Figure~\ref{fig: gr_mean} indicates that $\dot{M}_a$ significantly declines at $z<2$, but $\dot{M}_m$ shows more minor redshift evolution, with a small decline at $z<1$. Therefore, the fractional contribution from the merger channel to the total SMBH growth becomes larger as redshift decreases. Figure~\ref{fig: gr_mean} also indicates that the accretion channel always dominates over the merger channel at any redshift for the $p_\mathrm{merge}$ in \citet{Tremmel18}. In the upper-limit case where $p_\mathrm{merge}=1$, though, it is possible that the merger channel dominates over the accretion channel at $z<0.5$. \citet{Ricarte18} adopted a constant $p_\mathrm{merge}=0.1$, which is more physically plausible compared to $p_\mathrm{merge}=1$, and the corresponding $\dot{M}_m$ is 1~dex lower than the curve for $p_\mathrm{merge}=1$, leading to a $\dot{M}_m$ similar to ours based on \citet{Tremmel18} at $z<0.5$ but being smaller at higher redshifts. Therefore, at least at $z\gtrsim0.5$, the merger channel is negligible for the whole SMBH population, and it is likely that the merger channel also does not dominate at lower redshifts unless $p_\mathrm{merge}$ is much higher than expected.\par

\begin{figure*}
\includegraphics[width=\hsize]{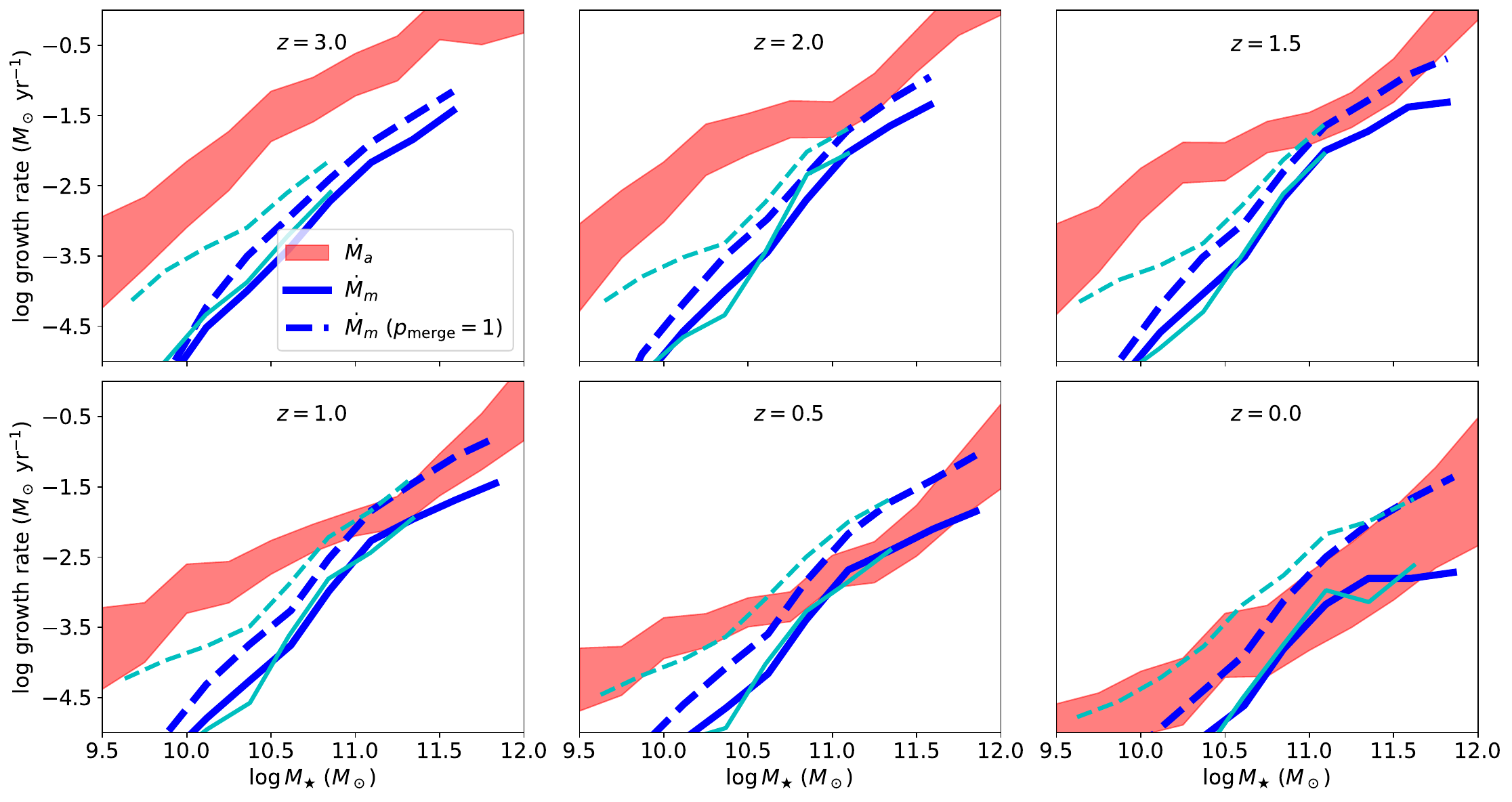}
\caption{The growth rates as functions of $M_\star$ at different redshifts. The red shaded regions represent the 90\% confidence range of $\dot{M}_a$ in \citet{Zou24}. The blue curves represent $\dot{M}_m$, and the corresponding solid and dashed curves represent the $p_\mathrm{merge}$ in \citet{Tremmel18} and $p_\mathrm{merge}=1$, respectively. The cyan curves correspond to TNG100 $\dot{M}_m$ when considering MBHs in dwarf galaxies, where the solid and dashed ones represent our best-guess and extreme, hard-limit cases, respectively.}
\label{fig: gr_mstar}
\end{figure*}

\begin{figure*}
\includegraphics[width=\hsize]{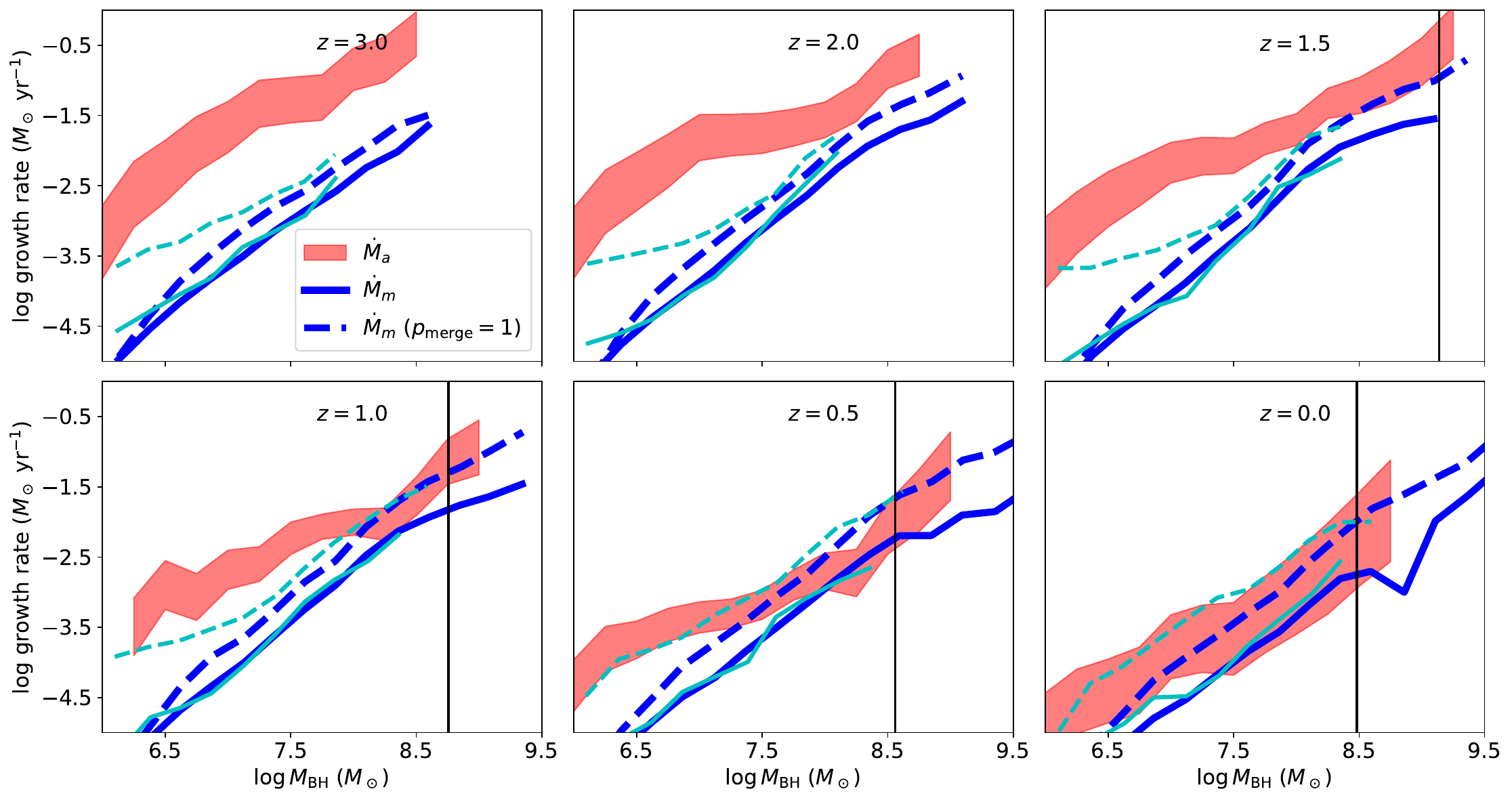}
\caption{Same as Figure~\ref{fig: gr_mstar}, but for the growth rates as functions of $M_\mathrm{BH}$, where we use our $M_\mathrm{BH}-M_\star$ scaling relation to convert $M_\mathrm{BH}$ to $M_\star$. The black vertical lines are the transitioning $M_\mathrm{BH}$ predicted in \citet{Pacucci20} when $\dot{M}_a=\dot{M}_m$ at the corresponding redshifts.}
\label{fig: gr_mbh}
\end{figure*}

We further measure $\dot{M}_m$ in $M_\star$ and $M_\mathrm{BH}$ bins at different redshifts, where we bin $M_\star$ or $M_\mathrm{BH}$ with a width of 0.25~dex and only keep bins with at least 50 sources. The results are tabulated in Tables~\ref{tbl: growthrate_mstar} and \ref{tbl: growthrate_mbh}, and $\dot{M}_a$ can be retrieved from Section~3.3 of \citet{Zou24}. Figures~\ref{fig: gr_mstar} and \ref{fig: gr_mbh} present $\dot{M}_m$ versus $M_\star$ and $M_\mathrm{BH}$. We plot the 90\% confidence range of $\dot{M}_a(M_\star, z)$ in \citet{Zou24} and use the $M_\mathrm{BH}-M_\star$ scaling relation in Section~\ref{sec: scaling} to convert $\dot{M}_a$ as a function of $(M_\star, z)$ to a function of $(M_\mathrm{BH}, z)$ in Figure~\ref{fig: gr_mbh}. To suppress $\dot{M}_m$ fluctuations, we directly calculate the expected merged $M_\mathrm{BH}$ for each galaxy merger event as $p_\mathrm{merge}M_\mathrm{BH}^\mathrm{secondary}$, where $M_\mathrm{BH}^\mathrm{secondary}$ is the mass of the secondary SMBH, to avoid introducing additional uncertainties from the random determination of whether two SMBHs would eventually merge. We also calculate $\dot{M}_m$ within $\pm3$ snapshots (i.e., spanning 0.9~Gyrs) and adopt their average value as $\dot{M}_m$ in each bin (i.e., we use \textit{sliding window} smoothing). In Figure~\ref{fig: gr_mbh}, we also explicitly mark the predicted transitioning $M_\mathrm{BH}$ where $\dot{M}_a=\dot{M}_m$ from \citet{Pacucci20} for comparison. The figures show that both $\dot{M}_a$ and $\dot{M}_m$ increase with $M_\star$ and $M_\mathrm{BH}$, and $\dot{M}_m$ gradually approaches $\dot{M}_a$ with rising mass and decreasing redshift. The solid blue $\dot{M}_m$ curves begin to overlap with the $\dot{M}_a$ shaded regions at $z<1$ when $M_\star\gtrsim10^{10.7}-10^{11.2}~M_\odot$ or $M_\mathrm{BH}\gtrsim10^{7.5}-10^{8.5}~M_\odot$, where the exact transitioning masses move to lower masses with decreasing redshift. If we set $p_\mathrm{merge}=1$, $\dot{M}_m$ would become comparable with $\dot{M}_a$ earlier at $z=1.5$ at the high-mass end ($M_\star\gtrsim10^{11}~M_\odot$ and $M_\mathrm{BH}\gtrsim10^{8.5}~M_\odot$), and the transitioning mass at lower redshift would decrease by $\approx0.3$~dex. Under this case, it is even possible that $\dot{M}_m(z=0)$ is comparable to $\dot{M}_a(z=0)$ for the whole mass range we are considering, although this is partly due to the elevated statistical uncertainties of $\dot{M}_a(z=0)$.\par

\begin{table*}
\caption{Merger-driven Growth Rates in $M_\star$ bins}
\label{tbl: growthrate_mstar}
\centering
\resizebox{\textwidth}{!}{
\begin{threeparttable}
\begin{tabular}{c|ccccccc}
\hline
\hline
$\log M_\star$ & $z=0$ & $z=0.5$ & $z=1$ & $z=1.5$ & $z=2$ & $z=2.5$ & $z=3$\\
\hline
$9.50-9.75$ & $-5.62~(-5.72)$ & $-5.77~(-5.63)$ & $-5.90~(-5.62)$ & $-6.59~(-6.40)$ & $-6.40~(-6.05)$ & $-6.80~(-6.19)$ & $-7.09~(-6.74)$\\
& $-6.25~(-5.90)$ & $-6.24~(-5.87)$ & $-6.14~(-5.84)$ & $-6.22~(-5.86)$ & $-6.27~(-6.06)$ & $-6.47~(-6.20)$ & $-6.87~(-6.45)$\\
$9.75-10.00$ & $-5.44~(-5.42)$ & $-5.63~(-5.18)$ & $-5.39~(-4.90)$ & $-5.42~(-4.95)$ & $-5.42~(-5.07)$ & $-5.45~(-5.02)$ & $-5.67~(-5.07)$\\
& $-6.08~(-5.41)$ & $-5.58~(-5.20)$ & $-5.39~(-5.04)$ & $-5.29~(-5.00)$ & $-5.27~(-4.90)$ & $-5.32~(-5.05)$ & $-5.36~(-5.25)$\\
$10.00-10.25$ & $-5.55~(-4.77)$ & $-5.14~(-4.56)$ & $-4.78~(-4.28)$ & $-4.60~(-4.06)$ & $-4.57~(-4.12)$ & $-4.59~(-4.17)$ & $-4.57~(-4.15)$\\
& $-5.47~(-4.95)$ & $-5.10~(-4.59)$ & $-4.80~(-4.31)$ & $-4.59~(-4.21)$ & $-4.57~(-4.18)$ & $-4.52~(-4.19)$ & $-4.52~(-4.23)$\\
$10.25-10.50$ & $-5.20~(-4.52)$ & $-4.87~(-4.13)$ & $-4.38~(-3.73)$ & $-4.14~(-3.59)$ & $-4.12~(-3.56)$ & $-4.02~(-3.54)$ & $-3.90~(-3.51)$\\
& $-5.12~(-4.39)$ & $-4.65~(-4.07)$ & $-4.27~(-3.75)$ & $-4.04~(-3.52)$ & $-4.00~(-3.51)$ & $-4.01~(-3.52)$ & $-4.00~(-3.50)$\\
$10.50-10.75$ & $-4.99~(-4.02)$ & $-3.91~(-3.43)$ & $-3.56~(-3.08)$ & $-3.39~(-2.91)$ & $-3.33~(-2.86)$ & $-3.28~(-2.75)$ & $-3.17~(-2.70)$\\
& $-4.63~(-3.92)$ & $-4.16~(-3.60)$ & $-3.75~(-3.26)$ & $-3.51~(-3.02)$ & $-3.45~(-2.97)$ & $-3.44~(-2.98)$ & $-3.40~(-2.95)$\\
$10.75-11.00$ & $-3.68~(-3.05)$ & $-3.23~(-2.64)$ & $-2.77~(-2.27)$ & $-2.60~(-2.19)$ & $-2.37~(-2.07)$ & $-2.63~(-2.20)$ & $-2.59~(-2.15)$\\
& $-3.85~(-3.16)$ & $-3.40~(-2.85)$ & $-2.98~(-2.52)$ & $-2.68~(-2.31)$ & $-2.70~(-2.35)$ & $-2.72~(-2.39)$ & $-2.72~(-2.40)$\\
$11.00-11.25$ & $-3.14~(-2.37)$ & $-2.82~(-2.10)$ & $-2.43~(-1.89)$ & $-2.01~(-1.65)$ & $-2.04~(-1.74)$ & $-2.26~(-1.93)$ & ---\\
& $-3.21~(-2.51)$ & $-2.69~(-2.17)$ & $-2.26~(-1.84)$ & $-2.00~(-1.65)$ & $-2.03~(-1.71)$ & $-2.07~(-1.79)$ & $-2.17~(-1.88)$\\
$11.25-11.50$ & $-1.73~(-1.77)$ & $-2.47~(-1.74)$ & $-1.92~(-1.40)$ & --- & --- & --- & ---\\
& $-2.83~(-2.01)$ & $-2.41~(-1.72)$ & $-1.96~(-1.45)$ & $-1.72~(-1.28)$ & $-1.65~(-1.29)$ & $-1.74~(-1.38)$ & $-1.84~(-1.51)$\\
$11.50-11.75$ & $-2.55~(-1.39)$ & --- & --- & --- & --- & --- & ---\\
& $-2.50~(-1.61)$ & $-2.10~(-1.40)$ & $-1.69~(-1.07)$ & $-1.38~(-0.92)$ & $-1.35~(-0.95)$ & $-1.40~(-1.06)$ & $-1.43~(-1.15)$\\
$11.75-12.00$ & --- & --- & --- & --- & --- & --- & ---\\
& $-2.98~(-1.41)$ & $-1.85~(-1.05)$ & $-1.45~(-0.80)$ & $-1.30~(-0.69)$ & --- & --- & ---\\
\hline
\hline
\end{tabular}
\begin{tablenotes}
\item
\textit{Notes.} The first column lists the $\log M_\star$ range of each bin, and the other columns show $\log\dot{M}_m$ in $M_\odot~\mathrm{yr^{-1}}$ at the corresponding bins. Within each cell, the top and bottom parts show the TNG100 and TNG300 results, respectively. The values in the parentheses correspond to $p_\mathrm{merge}=1$ and are thus upper limits.
\end{tablenotes}
\end{threeparttable}
}
\end{table*}

\begin{table*}
\caption{Merger-driven Growth Rates in $M_\mathrm{BH}$ bins}
\label{tbl: growthrate_mbh}
\centering
\resizebox{\textwidth}{!}{
\begin{threeparttable}
\begin{tabular}{c|ccccccc}
\hline
\hline
$\log M_\mathrm{BH}$ & $z=0$ & $z=0.5$ & $z=1$ & $z=1.5$ & $z=2$ & $z=2.5$ & $z=3$\\
\hline
$6.00-6.25$ & $-5.39~(-5.49)$ & $-5.88~(-5.63)$ & $-5.43~(-5.49)$ & $-5.38~(-5.43)$ & $-5.20~(-5.33)$ & $-5.15~(-5.14)$ & $-5.00~(-4.93)$\\
& $-5.80~(-5.73)$ & $-5.69~(-5.59)$ & $-5.55~(-5.52)$ & $-5.43~(-5.36)$ & $-5.24~(-5.31)$ & $-5.09~(-5.07)$ & $-4.97~(-4.94)$\\
$6.25-6.50$ & $-5.38~(-5.07)$ & $-5.24~(-5.02)$ & $-5.14~(-4.91)$ & $-4.94~(-4.92)$ & $-5.00~(-4.55)$ & $-4.70~(-4.51)$ & $-4.59~(-4.33)$\\
& $-5.53~(-5.20)$ & $-5.25~(-4.98)$ & $-5.08~(-4.91)$ & $-4.92~(-4.77)$ & $-4.76~(-4.61)$ & $-4.62~(-4.45)$ & $-4.55~(-4.36)$\\
$6.50-6.75$ & $-5.10~(-4.63)$ & $-4.90~(-4.51)$ & $-4.57~(-4.34)$ & $-4.62~(-4.33)$ & $-4.42~(-4.19)$ & $-4.21~(-3.93)$ & $-4.07~(-3.85)$\\
& $-5.20~(-4.73)$ & $-4.85~(-4.51)$ & $-4.68~(-4.36)$ & $-4.53~(-4.25)$ & $-4.38~(-4.07)$ & $-4.25~(-3.95)$ & $-4.16~(-3.86)$\\
$6.75-7.00$ & $-5.02~(-4.25)$ & $-4.75~(-4.20)$ & $-4.48~(-3.94)$ & $-4.25~(-3.78)$ & $-4.11~(-3.75)$ & $-4.03~(-3.61)$ & $-4.11~(-3.40)$\\
& $-4.81~(-4.25)$ & $-4.49~(-4.06)$ & $-4.34~(-3.92)$ & $-4.21~(-3.86)$ & $-4.06~(-3.68)$ & $-3.93~(-3.54)$ & $-3.83~(-3.48)$\\
$7.00-7.25$ & $-4.86~(-4.47)$ & $-4.16~(-3.69)$ & $-4.11~(-3.64)$ & $-3.93~(-3.48)$ & $-3.71~(-3.42)$ & $-3.64~(-3.24)$ & $-3.43~(-3.15)$\\
& $-4.55~(-3.90)$ & $-4.22~(-3.73)$ & $-4.00~(-3.65)$ & $-3.88~(-3.48)$ & $-3.71~(-3.32)$ & $-3.57~(-3.17)$ & $-3.51~(-3.12)$\\
$7.25-7.50$ & $-4.46~(-3.71)$ & $-3.87~(-3.37)$ & $-3.63~(-3.27)$ & $-3.51~(-3.25)$ & $-3.30~(-2.92)$ & $-3.19~(-2.73)$ & $-3.23~(-2.65)$\\
& $-4.22~(-3.63)$ & $-3.85~(-3.42)$ & $-3.62~(-3.26)$ & $-3.49~(-3.15)$ & $-3.31~(-2.97)$ & $-3.20~(-2.87)$ & $-3.15~(-2.81)$\\
$7.50-7.75$ & $-3.65~(-3.31)$ & $-3.66~(-3.12)$ & $-3.15~(-2.75)$ & $-3.10~(-2.75)$ & $-2.97~(-2.63)$ & $-2.91~(-2.51)$ & $-2.59~(-2.45)$\\
& $-3.80~(-3.31)$ & $-3.51~(-3.06)$ & $-3.25~(-2.86)$ & $-3.10~(-2.79)$ & $-2.97~(-2.66)$ & $-2.89~(-2.60)$ & $-2.86~(-2.57)$\\
$7.75-8.00$ & $-3.11~(-2.72)$ & $-3.13~(-2.54)$ & $-2.74~(-2.34)$ & $-2.53~(-2.34)$ & $-2.44~(-2.23)$ & $-2.65~(-2.44)$ & $-2.61~(-2.33)$\\
& $-3.60~(-3.01)$ & $-3.15~(-2.73)$ & $-2.90~(-2.56)$ & $-2.70~(-2.40)$ & $-2.63~(-2.33)$ & $-2.58~(-2.28)$ & $-2.58~(-2.25)$\\
$8.00-8.25$ & $-2.88~(-2.33)$ & $-2.62~(-2.15)$ & $-2.50~(-2.03)$ & $-2.27~(-1.73)$ & $-2.07~(-1.73)$ & $-2.43~(-1.87)$ & ---\\
& $-3.26~(-2.58)$ & $-2.81~(-2.32)$ & $-2.48~(-2.07)$ & $-2.27~(-1.90)$ & $-2.25~(-1.93)$ & $-2.26~(-1.97)$ & $-2.24~(-1.96)$\\
$8.25-8.50$ & $-2.61~(-1.87)$ & $-2.61~(-1.93)$ & $-2.52~(-1.73)$ & $-2.13~(-1.74)$ & --- & --- & ---\\
& $-2.78~(-2.19)$ & $-2.47~(-1.92)$ & $-2.13~(-1.70)$ & $-1.95~(-1.60)$ & $-1.93~(-1.58)$ & $-1.99~(-1.64)$ & $-2.01~(-1.64)$\\
$8.50-8.75$ & --- & --- & --- & --- & --- & --- & ---\\
& $-2.52~(-1.72)$ & $-2.20~(-1.62)$ & $-1.93~(-1.43)$ & $-1.77~(-1.34)$ & $-1.70~(-1.34)$ & $-1.72~(-1.37)$ & $-1.65~(-1.49)$\\
$8.75-9.00$ & --- & --- & --- & --- & --- & --- & ---\\
& $-3.07~(-1.59)$ & $-2.20~(-1.43)$ & $-1.76~(-1.22)$ & $-1.63~(-1.13)$ & $-1.57~(-1.18)$ & $-1.31~(-1.09)$ & ---\\
$9.00-9.25$ & --- & --- & --- & --- & --- & --- & ---\\
& $-2.00~(-1.44)$ & $-1.90~(-1.12)$ & $-1.64~(-0.99)$ & $-1.54~(-1.01)$ & $-1.30~(-0.93)$ & --- & ---\\
$9.25-9.50$ & --- & --- & --- & --- & --- & --- & ---\\
& $-1.63~(-1.22)$ & $-1.85~(-1.01)$ & $-1.46~(-0.72)$ & --- & --- & --- & ---\\
$9.50-9.75$ & --- & --- & --- & --- & --- & --- & ---\\
& $-1.23~(-1.01)$ & $-1.55~(-0.77)$ & --- & --- & --- & --- & ---\\
\hline
\hline
\end{tabular}
\begin{tablenotes}
\item
\textit{Notes.} Same as Table~\ref{tbl: growthrate_mstar}, but for $\dot{M}_m$ in $M_\mathrm{BH}$ bins.
\end{tablenotes}
\end{threeparttable}
}
\end{table*}

In the analyses above, we did not consider merged masses from MBHs in dwarf galaxies, which could cause underestimation of $\dot{M}_m$. We first estimate our best-guess contributions from MBHs in dwarf galaxies by extending the MBH seeding down to $M_\star=10^8~M_\odot$, extrapolating the $p_\mathrm{merge}$ in \citet{Tremmel18} down to the dwarf-galaxy regime, and setting $\dot{M}_a(M_\star<10^{9.5}~M_\odot)=\dot{M}_a(M_\star=10^{9.5}~M_\odot)\times M_\star/(10^{9.5}~M_\odot)$. When seeding MBHs into dwarf galaxies, we also randomly determine if the galaxy contains a MBH seed with a Bernoulli realization, where the success probability is set to the MBH occupation fraction in Equation~1 of \citet{Gallo19}. We rerun our analyses for TNG100.\footnote{We present TNG100 results for this analysis because it has a better baryonic mass resolution suitable to probe the dwarf regime. Besides, our further results indicate that these added contributions from dwarf galaxies, even in the most extreme case, should have little impact on the high-mass end, and thus the analysis of TNG300 is unnecessary.} The resulting best-guess $\dot{M}_m$ is plotted in Figures~\ref{fig: gr_mstar} and \ref{fig: gr_mbh} and is similar to the original $\dot{M}_m$, indicating that the contributions from dwarf galaxies may be small. However, the above best-guess contribution from dwarf galaxies may not be reliable due to the highly limited understanding of MBHs in dwarf galaxies, as outlined in Section~\ref{sec: bhseed}, and we thus also provide hard upper limits. For this purpose, we set all the relevant factors to extreme values, i.e., $\dot{M}_a(M_\star<10^{9.5}~M_\odot)=\dot{M}_a(M_\star=10^{9.5}~M_\odot)$, $p_\mathrm{merge}=1$, and the MBH occupation fraction in dwarf galaxies being 100\%. The $\dot{M}_m$ in this extreme, hard-limit case is also presented in Figures~\ref{fig: gr_mstar} and \ref{fig: gr_mbh}, and it elevates the original $\dot{M}_m(p_\mathrm{merge}=1)$ upper limit primarily at the low-mass end ($M_\mathrm{BH}\lesssim10^{10.5}~M_\odot$ and $M_\mathrm{BH}\lesssim10^{7.5}~M_\odot$). Since the low-mass $\dot{M}_m$ at $z\gtrsim1$ is far smaller than $\dot{M}_a$, this further elevated $\dot{M}_m$ is still below $\dot{M}_a$ and nearly does not change the overall SMBH growth at $z\gtrsim1$. However, at $z\lesssim0.5$, the elevation of $\dot{M}_m$ becomes more important and may cause $\dot{M}_m$ to dominate over $\dot{M}_a$ at $z=0$. Therefore, besides the regulation of $p_\mathrm{merge}$, there is additional room for $\dot{M}_m$ to reach higher values and become more important in the overall SMBH growth due to the possible contributions from MBHs in dwarf galaxies. This effect is negligible in our best-guess case, but, in principle, it is likely to cause more elevations in $\dot{M}_m$. In our extreme case, this effect matters mainly in the low-mass and low-$z$ regime. It should also be noted that this low-mass regime retains a high fraction of memory about our input seeds (Section~\ref{sec: massassembly}) and is thus sensitive to the seeding.\par
Overall, the accretion channel almost always dominates at $z>1$, where most of the growth happens (cf. Figure~\ref{fig: bhmf_evo}). At lower redshift, the merger-driven growth may become comparable with the accretion-driven growth, particularly at the high-mass end with $M_\star\gtrsim10^{11}~M_\odot$. The transitioning mass can be pushed down further, depending upon $p_\mathrm{merge}$ and contributions from MBHs in dwarf galaxies. This finding, however, differs from the original TNG-simulated results, where the TNG SMBH growth is instead dominated by mergers for galaxies with $M_\star\gtrsim10^{10.5}~M_\odot$ or $M_\mathrm{BH}\gtrsim10^{8.5}~M_\odot$ at $z<1-3$ \citep{Weinberger18}. The main reason is that the TNG SMBH accretion in galaxies with $M_\star\gtrsim10^{10.5}~M_\odot$ tends to be strongly suppressed due to very efficient feedback in TNG simulations, which can also be seen in Figure~\ref{fig: bhar_tng} where the TNG $\dot{M}_a$ does not increase with $M_\star$ at $M_\star\gtrsim10^{10.5}~M_\odot$. Since the TNG $\dot{M}_a$ is significantly below the observational measurements for massive galaxies at $z\lesssim2$ (Figure~\ref{fig: bhar_tng}), we argue that the accretion-driven growth should have more contribution than inferred from the original TNG simulations.\par
An observable metric that might indicate the relative contributions from different channels is the SMBH spin. Growth through coherent accretion (if applicable) can efficiently spin up SMBHs, while chaotic mergers tend to cause low-to-intermediate spins (e.g., \citealt{Reynolds19} and references therein). Only about two dozen SMBHs at $z\lesssim1$ have spin measurements (e.g., \citealt{Vasudevan16}), but they reveal the tendency that SMBHs with $M_\mathrm{BH}\lesssim10^{7.5}~M_\odot$ generally have high spins, and more massive SMBHs may have lower spins (see Figure~6 in \citealt{Vasudevan16}). Although large uncertainties and selection biases exist for the currently available spin measurements, this pattern is broadly consistent with our finding that SMBHs with $M_\mathrm{BH}\lesssim10^{7.5}~M_\odot$ should generally have accretion-driven growth.\par

% Mass assembly
\subsection{Mass-Assembly History}
\label{sec: massassembly}
We probe the redshift evolution of $M_\mathrm{BH}$ in this section. A single galaxy at $z=0$ usually has many high-redshift progenitors, and we adopt two kinds of definitions of the evolutionary track. First, we adopt the main progenitor branch of its merger tree, which picks out a single progenitor galaxy (if any) with the most massive history, as defined in \citet{DeLucia07}, in each relevant snapshot, as the \textit{main-progenitor} track. Galaxies on the main progenitor branch are usually but not necessarily the most massive ones at the corresponding snapshots. Second, we adopt the total mass of all the progenitors at each snapshot as the \textit{all-progenitor} track. As we will show below, different definitions can lead to different mass-assembly behaviors.\par
\begin{figure*}
\includegraphics[width=\hsize]{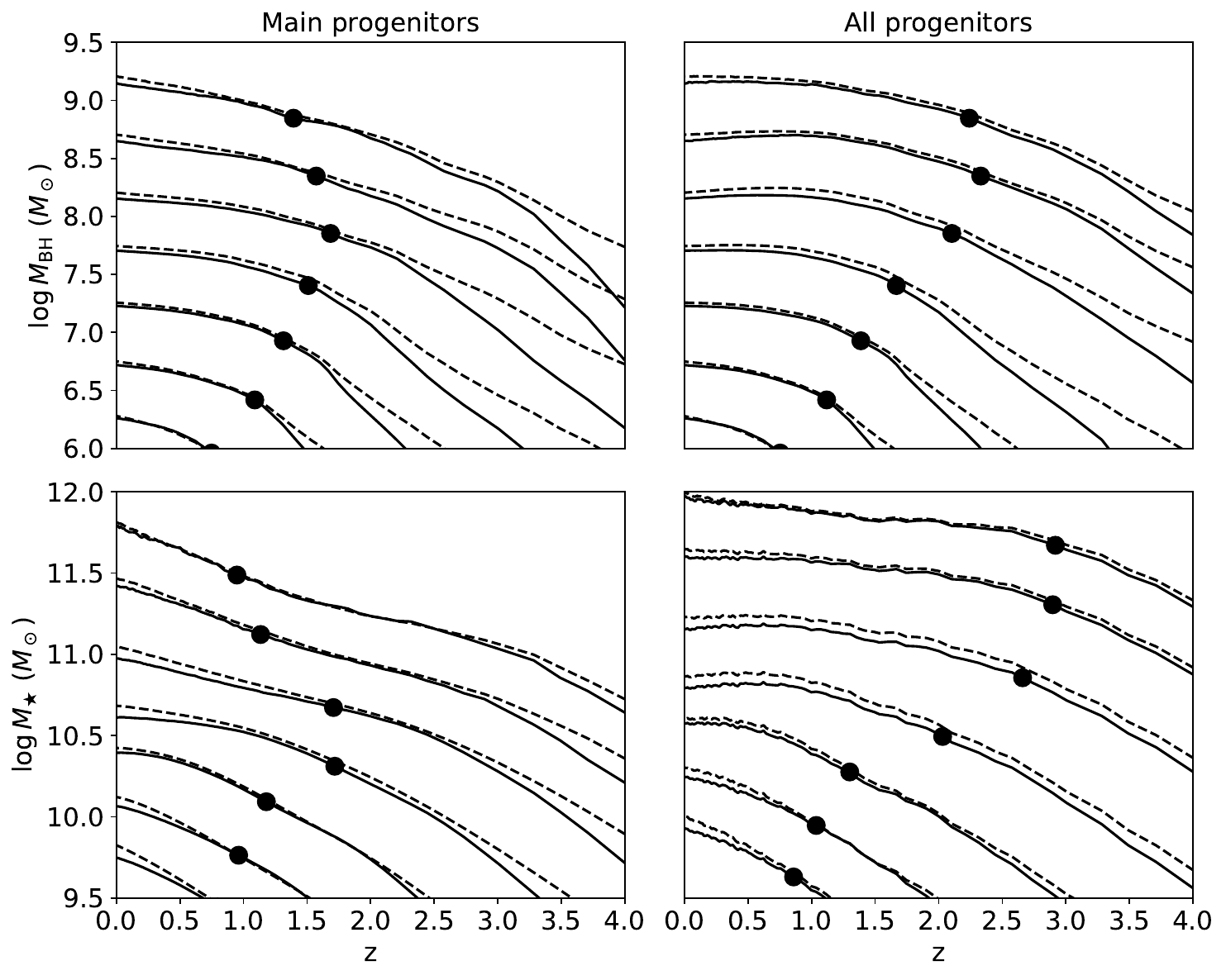}
\caption{The evolutionary tracks of $M_\mathrm{BH}$ (top) and $M_\star$ (bottom) for the main progenitors' masses (left) and all the progenitors' total masses (right), where sources are binned with a $M_\mathrm{BH}(z=0)$ width of 0.5~dex. The solid and dashed curves are $\log[\mathrm{median}(M_\mathrm{BH})]$ ($\log[\mathrm{median}(M_\star)]$) and $\log[\mathrm{mean}(M_\mathrm{BH})]$ ($\log[\mathrm{mean}(M_\star)]$) tracks, respectively. The large points on the curves represent when half of the masses are assembled.}
\label{fig: mbhevo}
\end{figure*}

We bin our sources with a $M_\mathrm{BH}(z=0)$ width of 0.5~dex and plot their main-progenitor and all-progenitor evolutionary tracks of $M_\mathrm{BH}$ and $M_\star$ in Figure~\ref{fig: mbhevo}. We plot both $\log[\mathrm{median}(M_\mathrm{BH})]$ ($\log[\mathrm{median}(M_\star)]$) and $\log[\mathrm{mean}(M_\mathrm{BH})]$ ($\log[\mathrm{mean}(M_\star)]$) tracks. These two types of definitions can lead to different main-progenitor tracks at early stages because the mass distribution is broad and unsymmetrical -- for example, many galaxies may not even have SMBHs seeded yet (i.e., $M_\mathrm{BH}=0$) as their $M_\star$ values have not reached our threshold of $10^{9.5}~M_\odot$. However, the late-stage evolutionary tracks become largely insensitive to the definition. The figure indicates that half of $M_\mathrm{BH}$ is generally assembled by $z\approx0.8-2$, consistent with the findings in \citet{Shankar20a}. Besides, smaller SMBHs generally reach their half $M_\mathrm{BH}$ at lower redshifts for the all-progenitor tracks, which is referred to as the \textit{cosmic downsizing}, and their host galaxies have the same behavior. Regarding the main-progenitor tracks, massive galaxies with $M_\star(z=0)\gtrsim10^{11}~M_\odot$ and their SMBHs instead show an \textit{upsizing} phenomenon. This difference between the main-progenitor and all-progenitor tracks is inherent in the hierarchical formation, as discussed in detail in Sections 3 and 4 of \citet{Neistein06}.\par

\begin{figure}
\includegraphics[width=\hsize]{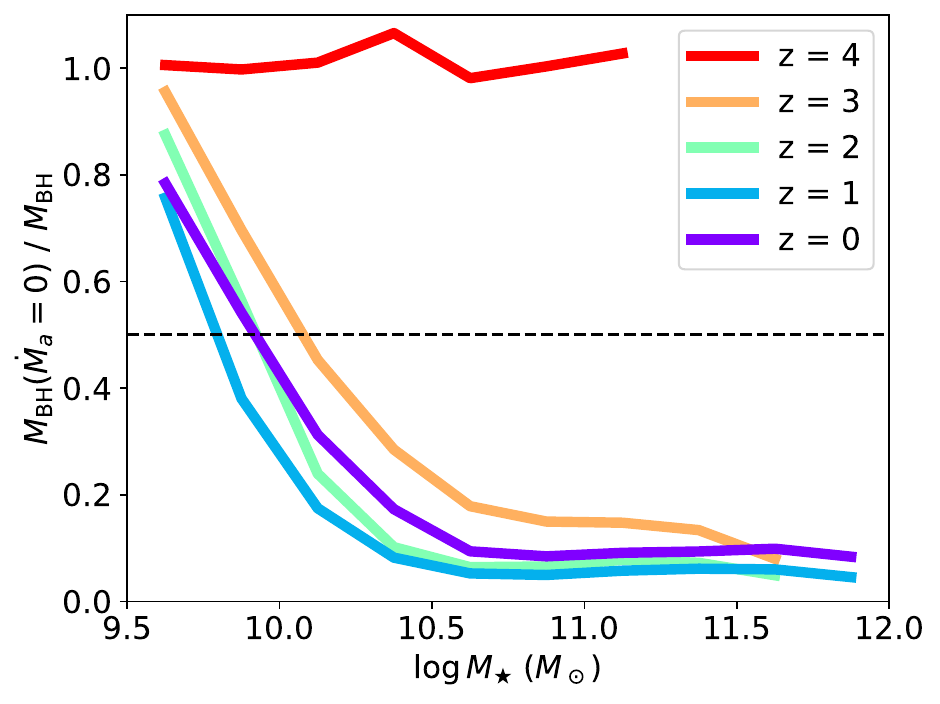}
\caption{Median $M_\mathrm{BH}(\dot{M}_a=0)/M_\mathrm{BH}$ versus $M_\star$ at different redshifts. The horizontal black dashed line represents $M_\mathrm{BH}(\dot{M}_a=0)/M_\mathrm{BH}=0.5$.}
\label{fig: seedmemory}
\end{figure}

Figure~\ref{fig: mbhevo} indicates that the final $M_\mathrm{BH}$ is generally several orders of magnitude higher than the initially seeded $M_\mathrm{BH}$, and thus the final $M_\mathrm{BH}$ is dominated by the accreted mass at $z<4$. To further assess this point quantitatively, we rerun our analyses in Section~\ref{sec: methodology} but set $\dot{M}_a=0$. The ratio between the resulting $M_\mathrm{BH}(\dot{M}_a=0)$ and our original $M_\mathrm{BH}$ represents the memory the SMBH records about its SMBH seed, where a smaller ratio indicates that more memory is lost. Note that $M_\mathrm{BH}$ can be contributed by multiple progenitors, and thus the seeding cannot be represented by a single SMBH seed; instead, we should follow the whole merger tree and use $M_\mathrm{BH}(\dot{M}_a=0)$ to quantify the cumulative contribution from all the relevant SMBH seeds to the final $M_\mathrm{BH}$. We plot $M_\mathrm{BH}(\dot{M}_a=0)/M_\mathrm{BH}$ as a function of $M_\star$ at different redshifts in Figure~\ref{fig: seedmemory}. The $z=4$ curve, by construction, is roughly unity. The $z=3$ curve rapidly decreases with $M_\star$ and is below $M_\mathrm{BH}(\dot{M}_a=0)/M_\mathrm{BH}=0.5$ when $M_\star>10^{10.1}~M_\odot$. The transitioning $M_\star$ corresponding to $M_\mathrm{BH}(\dot{M}_a=0)/M_\mathrm{BH}=0.5$ only slightly evolves with redshift after $z=3$, partly because we must keep seeding SMBHs into newly formed massive galaxies. It is also worth noting that $M_\mathrm{BH}(\dot{M}_a=0)/M_\mathrm{BH}$ is generally above 0.5 when $M_\star<10^{10}~M_\odot$, indicating that our results for these low-mass galaxies are largely determined by the $M_\mathrm{BH}$ seeding choice. This phenomenon has an important implication that only MBHs in low-mass galaxies with $M_\star<10^{10}~M_\odot$ record remnant information about their seeds, which has served as one pillar of the science related to MBHs in low-mass galaxies (e.g., \citealt{Greene20} and references therein). Overall, Figure~\ref{fig: seedmemory} indicates that the seeding memory is small when $M_\star\gtrsim10^{10}~M_\odot$ and $z\lesssim3$, and this is the regime where we regard this work to provide genuine physical insights into the SMBH population. We do note, though, the seeding memory would be elevated if we adopt a much higher seeding scaling relation, and Section~\ref{sec: differentseed} will discuss this in more detail.

% Wandering
\subsection{Wandering SMBHs}
\label{sec: wander}
Given the non-unity $p_\mathrm{merge}$, some SMBHs cannot reach the final coalescence after galaxy mergers and thus become long-lived wanderers. The wandering population may constitute a nonnegligible fraction of the total mass budget, and thus the classical Soltan argument needs corrections to account for wandering SMBHs \citep{Kulier15}. Besides, wandering SMBHs may have important observational implications and can sometimes shine as hyperluminous \mbox{X-ray} sources, dual AGNs, or off-nucleus tidal disruption events (e.g., \citealt{Ricarte21a}). This population is also under active theoretical research, and we briefly present our predictions based on our simplified $p_\mathrm{merge}$ method in this section.\par
\begin{figure}
\includegraphics[width=\hsize]{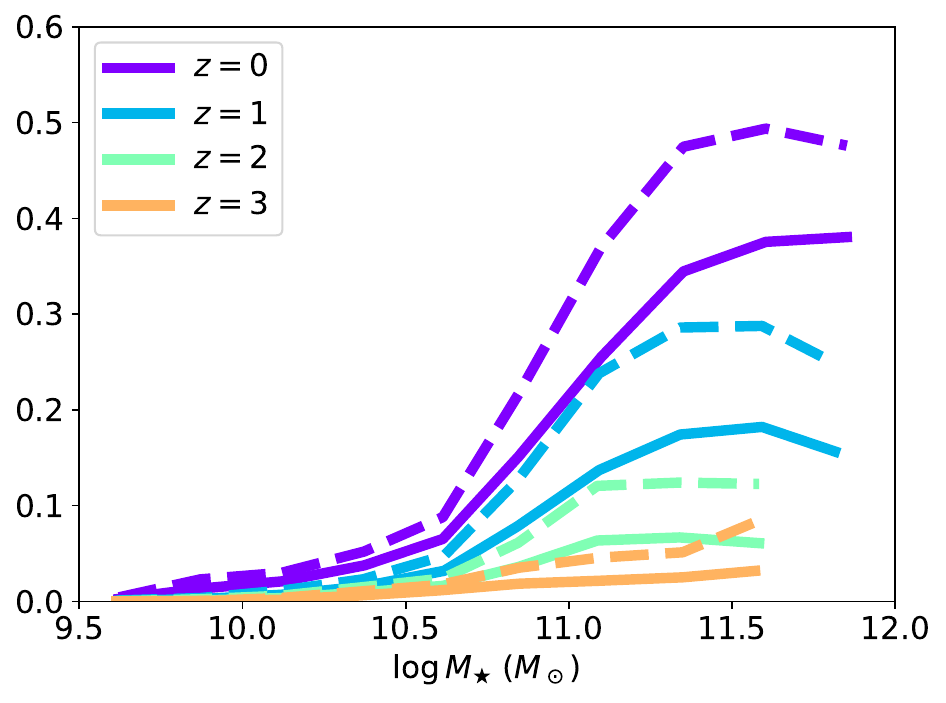}
\caption{The mass fraction of wandering SMBHs among all the SMBHs versus $\log M_\star$ at different redshifts. The solid and dashed curves correspond to the $p_\mathrm{merge}$ in \citet{Tremmel18} and $p_\mathrm{merge}=0$, respectively. The dashed curves represent the upper limits.}
\label{fig: wandermassfrac}
\end{figure}

We first show the fraction of $M_\mathrm{BH}$ locked in wandering SMBHs among all the SMBHs (i.e., central + wandering) in Figure~\ref{fig: wandermassfrac}, where we present both the results based on the $p_\mathrm{merge}$ in \citet{Tremmel18} and $p_\mathrm{merge}=0$, with the latter representing upper limits. The fraction increases with $M_\star$ and decreases with redshift. It is generally small ($<5\%$) when $M_\star<10^{10.5}~M_\odot$, but it increases to over 30\% for massive galaxies with $M_\star=10^{11.5}~M_\odot$ at $z=0$. Globally, 23.4\% ($<31.3\%$), 9.1\% ($<14.9\%$), 3.6\% ($<6.5\%$), and 1.6\% ($<3.1\%$) of SMBH mass is in wandering SMBHs at $z=0$, 1, 2, and 3, respectively, where the values in parentheses are the upper limits based on $p_\mathrm{merge}=0$. The $z=0$ numbers are roughly consistent with previous works in an order-of-magnitude sense. \citet{Kulier15} predicted the wandering mass fraction at $z=0$ to be 11.2\% on average and strongly increasing with rising $M_\star$ and cosmic density (i.e., void versus cluster). \citet{Ricarte21b} predicted the fraction to be $\approx10\%$, however, with little dependence on $M_\star$. Our value is higher possibly because our central SMBH growth is lower than in previous works (see Section~\ref{sec: scaling}). Although the mass fraction of wandering SMBHs is nonnegligible, it is generally challenging to find them through direct observations because they mostly remain dormant and can hardly grow through either accretion or mergers, but observational searches for them are still possible, as predicted in, e.g., \citet{Guo20} and \citet{Ricarte21a}.\par

\begin{figure}
\includegraphics[width=\hsize]{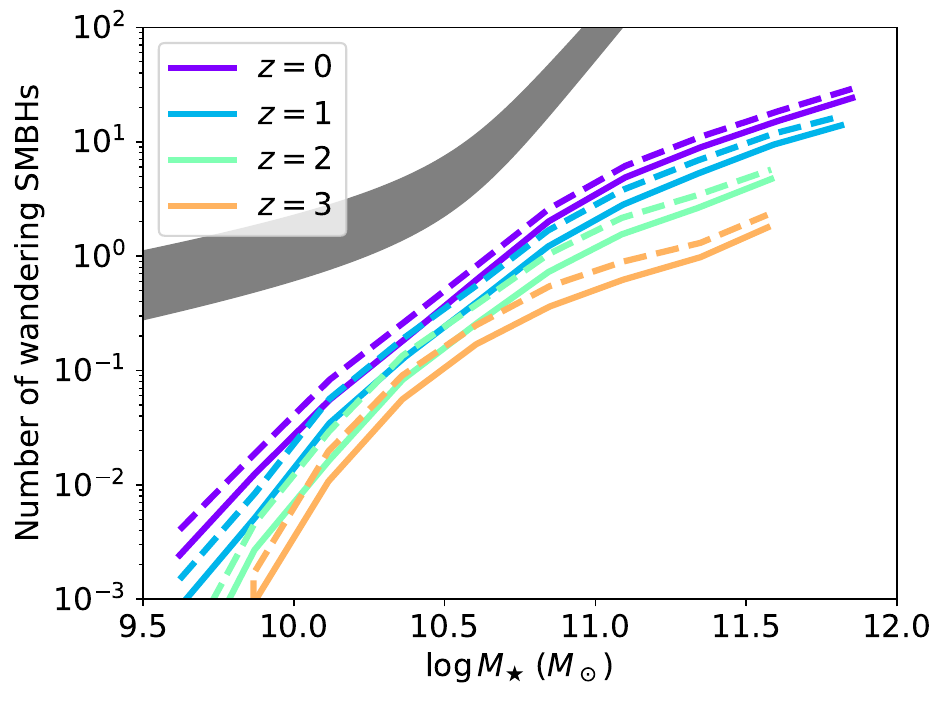}
\caption{The average number of wandering SMBHs per galaxy versus $M_\star$. The solid and dashed curves correspond to the $p_\mathrm{merge}$ in \citet{Tremmel18} and $p_\mathrm{merge}=0$, respectively. The gray shaded region is the relation in \citet{Ricarte21b} at $z=0$, where we convert the halo mass to $M_\star$ using the scaling relation in Section~3 of \citet{Moster13}. We underestimate the number of wandering MBHs because MBHs from dwarf galaxies are not considered in our analyses.}
\label{fig: nwander}
\end{figure}

There are some caveats worth noting. First, wandering SMBHs primarily originate from low-mass galaxies because low-mass galaxies have a higher number density than for massive galaxies, and minor mergers generally have small $p_\mathrm{merge}$. Therefore, most of these wandering SMBHs have $M_\mathrm{BH}$ close to our initially seeded $M_\mathrm{BH}$, and thus the wandering mass fraction is sensitive to our initial condition of SMBH seeds. Second, we neglected those wandering MBHs originating from dwarf galaxies or formed before $z=4$. \citet{DiMatteo23} showed that wandering MBHs may be dominated by less-massive ones, and \citet{Ricarte21b} showed that wandering MBHs may even dominate over central MBHs in terms of mass at $z=4$. Therefore, our lack of accounting for these MBHs may cause large underestimation of the wandering population. To illustrate this point, we plot the average number of wandering SMBHs per galaxy versus $M_\star$ in Figure~\ref{fig: nwander}, which shows a strong positive correlation, as expected from the hierarchical formation of galaxies. However, our number of wandering SMBHs is smaller than for \citet{Ricarte21b} by $\approx1-2$~dex, possibly primarily because we do not consider MBHs formed in dwarf galaxies. For example, \citet{Tremmel18b} predicted that MW-mass galaxies typically have roughly 12 wandering MBHs, but our predicted value is $<1$ because MW-like galaxies rarely undergo mergers with galaxies with $M_\star>10^{9.5}~M_\odot$ (see the left panel of Figure~\ref{fig: evotracks} for an example). Nevertheless, our wandering mass fraction (not the number counts) at $z=0$ is roughly similar to that in \citet{Ricarte21b}. Lastly, we reiterate that we still have little knowledge about wandering MBHs, either from observational or theoretical perspectives. \citet{vanDonkelaar24} showed that the properties of wandering MBHs, such as their numbers and dynamics, identified based on cosmological simulations strongly depend on the simulation resolution and sub-grid physics modeling MBHs. Therefore, caution should be taken when interpreting the relevant results.

% Different pmerge
\subsection{Impacts of the $p_\mathrm{merge}$ Choice}
\label{sec: differentpmerge}
\begin{figure}
\includegraphics[width=\hsize]{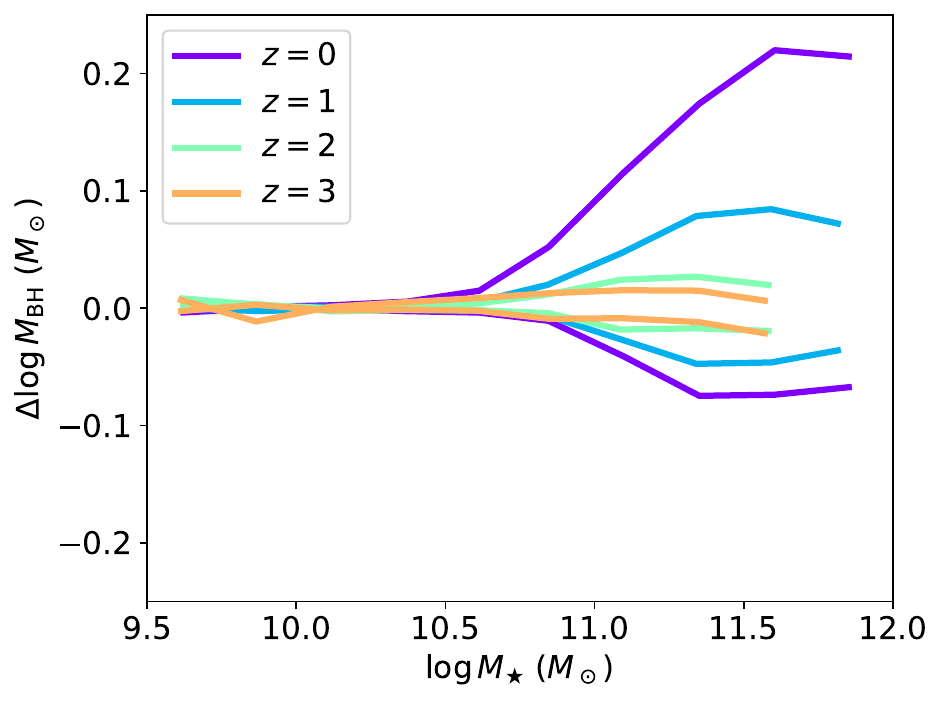}
\caption{Differences in $\log M_\mathrm{BH}$, denoted as $\Delta\log M_\mathrm{BH}$, when changing $p_\mathrm{merge}$ to 1 and 0. At each redshift, the upper and lower curves represent the running medians of $\log[M_\mathrm{BH}(p_\mathrm{merge}=1)/M_\mathrm{BH}]$ and $\log[M_\mathrm{BH}(p_\mathrm{merge}=0)/M_\mathrm{BH}]$, respectively.}
\label{fig: comp_mbh_pmerge}
\end{figure}

\begin{figure}
\includegraphics[width=\hsize]{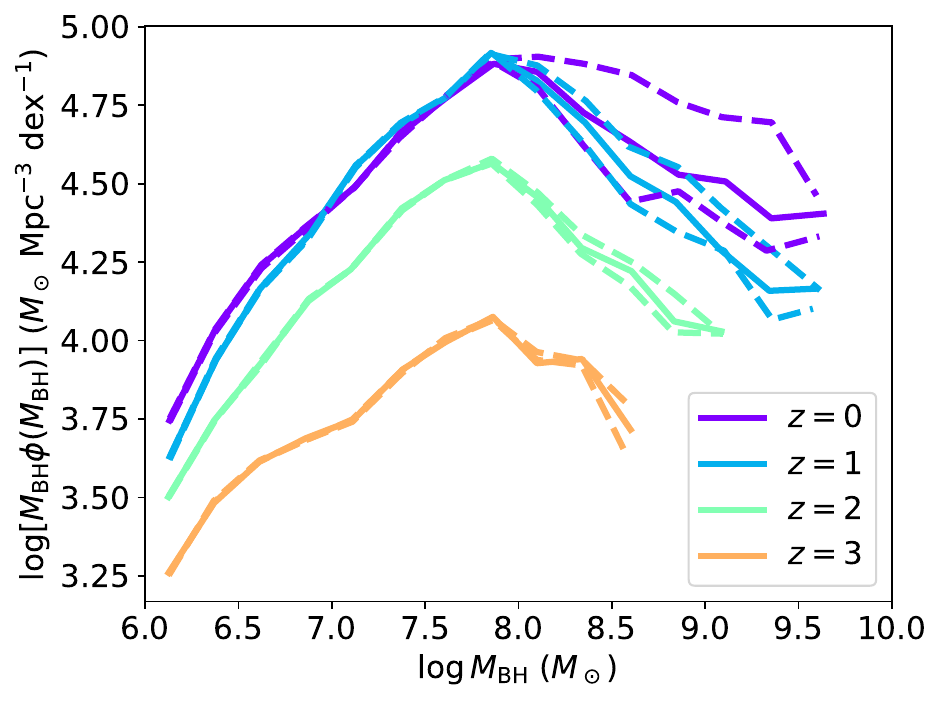}
\caption{The BHMFs corresponding to different $p_\mathrm{merge}$ choices. At each redshift, the solid curve is the original BHMF, while the upper and lower dashed curves represent $p_\mathrm{merge}=1$ and 0, respectively.}
\label{fig: comp_bhmf_pmerge}
\end{figure}

This section assesses the impact of the $p_\mathrm{merge}$ choice on our results. We choose two extreme cases of $p_\mathrm{merge}=1$ and $p_\mathrm{merge}=0$ and rerun the analyses in Section~\ref{sec: methodology}. We plot the differences between the resulting $\log M_\mathrm{BH}$ and our original $\log M_\mathrm{BH}$ versus $\log M_\star$ in Figure~\ref{fig: comp_mbh_pmerge}, which also represent the deviation from the scaling relations in Figure~\ref{fig: scaling}. Median $\Delta\log M_\mathrm{BH}$ increases with rising $M_\star$ and decreasing $z$. The deviation is almost negligible ($\lesssim0.03$~dex) across the whole $M_\star$ range at $z\geq2$ and the whole $z$ range at $M_\star\lesssim10^{11}~M_\odot$. $\Delta\log M_\mathrm{BH}$ reaches a maximum value of 0.2~dex for $p_\mathrm{merge}=1$ and a minimum value of $-0.1$~dex for $p_\mathrm{merge}=0$ at $M_\star>10^{11.5}~M_\odot$ and $z=0$, and such differences can hardly be the dominant element of the underlying systematic errors. We also present the BHMFs corresponding to different $p_\mathrm{merge}$ in Figure~\ref{fig: comp_bhmf_pmerge}. Similar to the findings in Figure~\ref{fig: comp_mbh_pmerge}, the BHMFs only have noticeable differences at $M_\mathrm{BH}>10^8~M_\odot$ and $z\leq1$, where the maximum deviations are 0.3~dex and $-0.2$~dex for $p_\mathrm{merge}=1$ and 0 at $z=0$, respectively. Integrating the BHMFs, we obtain $\rho_\mathrm{BH}=1.4\times10^5~M_\odot~\mathrm{Mpc}^{-3}$ and $1.9\times10^5~M_\odot~\mathrm{Mpc}^{-3}$ for $p_\mathrm{merge}=0$ and 1, respectively, which are similar to our original TNG300 $\rho_\mathrm{BH}$ ($1.5\times10^5~M_\odot~\mathrm{Mpc}^{-3}$; Section~\ref{sec: bhmf}). Therefore, although some differences in $M_\mathrm{BH}$ exist for massive galaxies at $z\leq1$, $p_\mathrm{merge}$ generally does not have major impacts on $M_\mathrm{BH}$ and the relevant quantities such as the BHMF (Section~\ref{sec: bhmf}) and the $M_\mathrm{BH}-M_\star$ relation (Section~\ref{sec: scaling}). This is expected because Section~\ref{sec: channelgrow} showed that the merger channel generally contributes little to the overall SMBH growth compared to the accretion channel, with possible exceptions at low redshifts, and $p_\mathrm{merge}$ only regulates the strength of the merger channel. In Section~\ref{sec: channelgrow}, we also tried elevating $\dot{M}_m$ upper limits by considering MBHs in dwarf galaxies and showed that the resulting $\dot{M}_m$ only differs from $\dot{M}_m(p_\mathrm{merge}=1)$ at the low-mass end where $\dot{M}_a$ usually dominates. Therefore, merged MBHs in dwarf galaxies almost do not impact $M_\mathrm{BH}$ (not explicitly plotted here).

However, $p_\mathrm{merge}$ plays an important role in results that are directly relevant to mergers. As has already been shown in Section~\ref{sec: channelgrow}, $\dot{M}_m$ strongly depends on $p_\mathrm{merge}$, and thus the exact transitioning $M_\mathrm{BH}$ and $M_\star$ values between the accretion-dominated ($\dot{M}_a>\dot{M}_m$) and merger-dominated ($\dot{M}_a<\dot{M}_m$) regimes depend upon $p_\mathrm{merge}$. Our results on wandering SMBHs in Section~\ref{sec: wander} also depend upon $p_\mathrm{merge}$. It should be noted that research on SMBH mergers and their implications is actively advancing and deserves dedicated further work (e.g., \citealt{Barausse21} and references therein). This work, however, does not aim to address $p_\mathrm{merge}$ but instead focuses on the overall central SMBH population, which is largely robust against the choice of $p_\mathrm{merge}$.

% Different seed
\subsection{Impacts of the Seeding Choice}
\label{sec: differentseed}
This section assesses the impacts of the initial seeding choice on our results. As mentioned in Section~\ref{sec: bhseed}, SMBHs are initially seeded following the AGN-based $M_\mathrm{BH}-M_\star$ scaling relation in \citet{Reines15}, which is lower than those based on dynamically measured $M_\mathrm{BH}$ by over one order of magnitude. We thus try adopting the latter one in \citet{Reines15}, $\log(M_\mathrm{BH}/M_\odot)=8.95+1.40[\log(M_\star/M_\odot)-11]$, as the seeding relation to examine how our results will change accordingly. We name it the \textit{high-seed} case in this subsection. We also try applying a redshift-dependent factor in \citet[see their Figure~3]{Pacucci24} to scale up our original seeding relation, named the \textit{zevo-seed} case (\textit{zevo} is the short for \textit{evolution with redshift}) in this subsection. In this case, the seeding relation would be elevated by 0.6, 1.0, 1.3, and 1.5 dex at $z=1$, 2, 3, and 4, respectively, compared to the local relation. Note that this redshift-dependent factor was not designed for $z<4$ as some of the assumptions in \citet{Pacucci24} would fail for non-early universe, and it is also not directly derived from observations. We solely use it to illustrate how a strongly redshift-dependent seeding relation can impact our results. We rerun our analyses with these two seeding choices while keeping all of our other settings the same as in Section~\ref{sec: methodology}.\par
\begin{figure}
\includegraphics[width=\hsize]{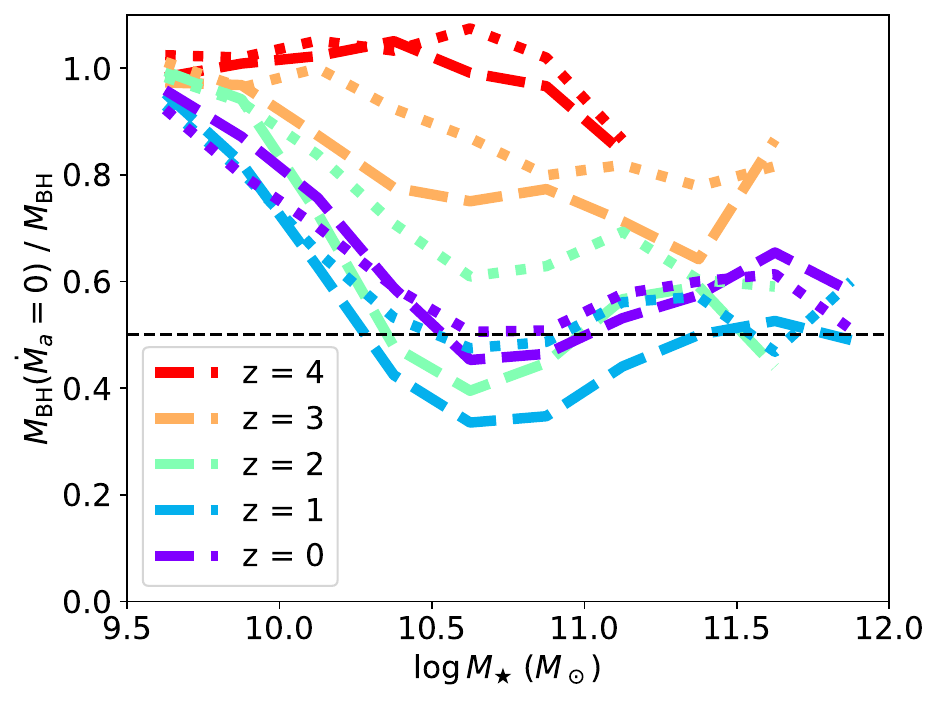}
\caption{Similar to Figure~\ref{fig: seedmemory}, but for different seeding choices. The dashed and dotted curves are for the high-seed and zevo-seed cases, respectively.}
\label{fig: seedmemory_diffseed}
\end{figure}

Before we present further scientific results, we first show memories about the seed masses, $M_\mathrm{BH}(\dot{M}_a=0)/M_\mathrm{BH}$, in Figure~\ref{fig: seedmemory_diffseed}. Unlike our original case in Figure~\ref{fig: seedmemory}, both the new high-seed and zevo-seed results retain $\gtrsim50\%$ of the memories about their seed masses, even in the high-mass regime, because the new seeds are much more massive than our previous ones. The fact that the median $M_\mathrm{BH}(\dot{M}_a=0)/M_\mathrm{BH}\approx0.5$ at $z<2$ in the high-mass regime indicates that the new seed masses are comparable to the accreted masses for massive galaxies. Since our original $M_\mathrm{BH}$ in the high-mass regime is primarily from the accreted masses (see Figure~\ref{fig: seedmemory}), increasing seed masses to the same level of the accreted masses would cause our $M_\mathrm{BH}$ in the high-mass regime to roughly double (i.e., increase by $\approx0.3$~dex).\par

\begin{figure*}
\includegraphics[width=\hsize]{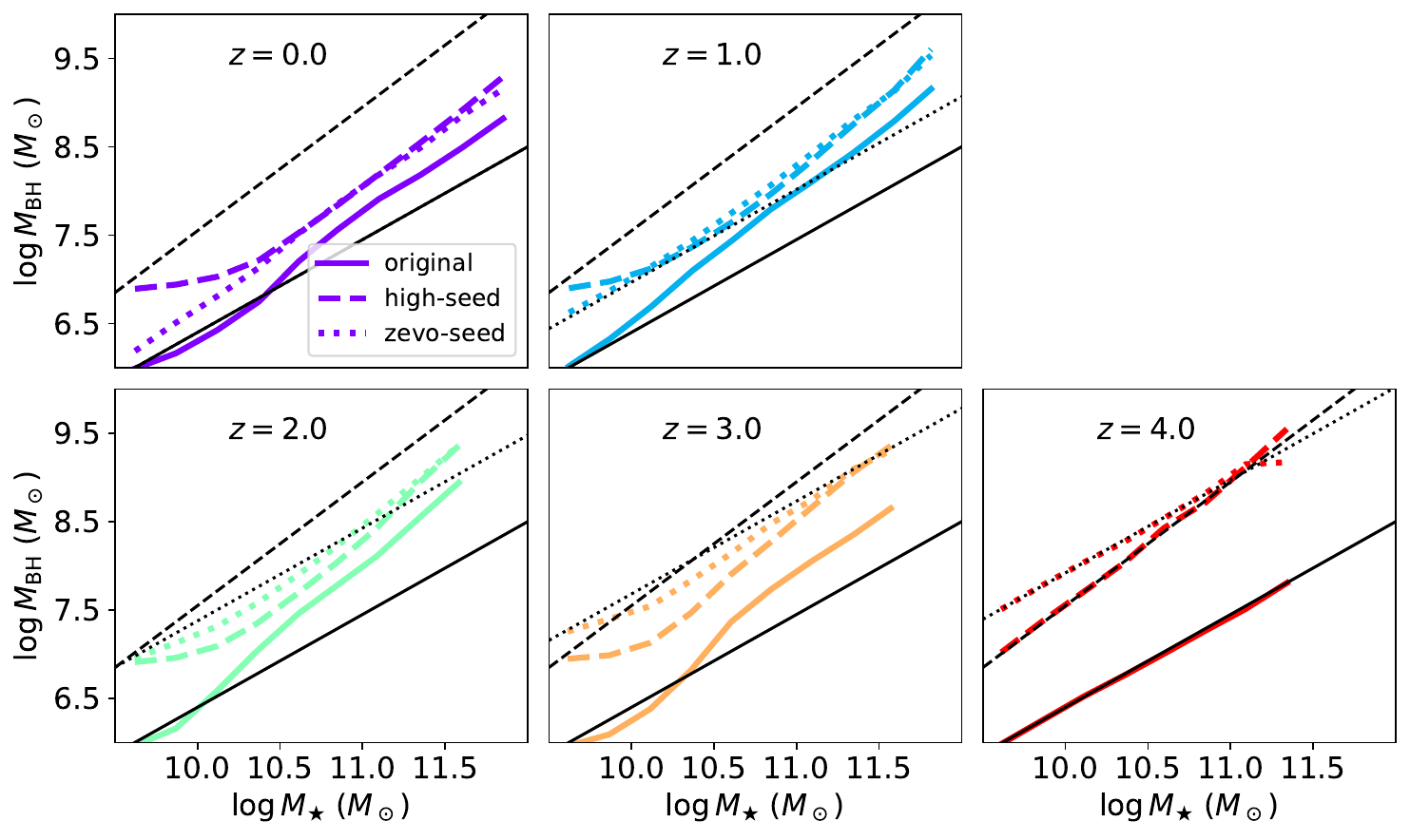}
\caption{Comparisons of the $M_\mathrm{BH}-M_\star$ scaling relations under different seeding choices. The black straight lines are our input seeding relations at the corresponding redshifts, and the colored curves are our recovered scaling relations. The dotted black line overlaps with the solid black line at $z=0$ and is thus invisible. The solid curves are our original ones following the AGN-based scaling relation in \citet{Reines15}, the dashed curves represent the high-seed case, and the dotted curves represent the zevo-seed case.}
\label{fig: scaling_diffseed}
\end{figure*}

With this conclusion in mind, we compare the resulting $M_\mathrm{BH}-M_\star$ scaling relations under different seeding choices in Figure~\ref{fig: scaling_diffseed}. In the high-seed case, our recovered scaling relation at $M_\star\gtrsim10^{10.5}~M_\odot$ is higher than our originally recovered one by over 1~dex at $z=4$, $\approx0.7$~dex at $z=3$, and $\approx0.3$~dex at $z\leq2$. At the low-mass end where the seed mass becomes increasingly dominant with decreasing $M_\star$, our recovered scaling relation flattens to connect to the input seeding relation in the smallest mass bin. Overall, as the Universe ages, the normalization of our recovered high-seed scaling relation keeps decreasing and becomes notably lower than the input one at $z=0$. This is because our accreted masses are unable to maintain this high normalization. However, this redshift dependence contradicts observations, where the scaling-relation normalization does not show apparent evolution with redshift (e.g., Figure~10 in \citealt{Li23}).\par

In the zevo-seed case, our recovered scaling relation is similar to the high-seed one at $M_\star\gtrsim10^{10.5}~M_\odot$ but does not become flat at the low-mass end. The seeding mainly happens at $z>3$ for the main progenitors of massive galaxies with $M_\star(z=0)\gtrsim10^{10.5}~M_\odot$ (see the bottom left panel of Figure~\ref{fig: mbhevo}, while less massive galaxies are seeded much later. Figure~\ref{fig: scaling_diffseed} shows that our input zevo-seed seeding relation (the dotted black lines) is similar to the high-seed one (the dashed black lines) at high redshifts and lowers down to our original one (the solid black lines) at low redshifts. Therefore, the impacts of the zevo-seed case on our restuls is effectively similar to the high-seed case for massive galaxies with $M_\star\gtrsim10^{10.5}~M_\odot$. With the redshift-dependent normalization, our recovered zevo-seed scaling relation is more consistent with the input one compared to the high-seed case and also increases with redshift, similar to the input. However, it should be noted that the primary reason for the redshift-dependence of our recovered relation is that the accretion at $z<4$ is not able to maintain the high $M_\mathrm{BH}-M_\star$ normalization at $z\approx4$, and SMBHs have to accumulate half of their current masses before $z\approx4$ for such a strongly redshift-dependent scaling relation to happen. This requires strong SMBH growth at $z>4$ and contradicts with previous works that the $M_\mathrm{BH}$ build-up is dominated by the $z<4$ epoch (e.g., \citealt{Vito16, Vito18, Shankar20a, Shen20, Habouzit21, Zhang23}). If true, such a strong SMBH growth at $z>4$ has to be hidden by obscuration to be missed by previous optical or \mbox{X-ray} surveys. Some JWST results indeed revealed stronger SMBH growth than our previous expectations (e.g., \citealt{Maiolino24, Pacucci23}), and future works are still needed to further constrain the SMBH growth in the early universe.\par
With $M_\mathrm{BH}$ elevated by $\gtrsim0.3$~dex, $\dot{M}_m$, which is proportional to the merging $M_\mathrm{BH}$, also increases by roughly the same factor. We found $\dot{M}_m<\dot{M}_a$ at $z\gtrsim2$ in both the high-seed and zevo-seed cases. At lower redshift, we found that $\dot{M}_m$ is coincidentally similar to our original $p_\mathrm{merge}=1$ curves, with slight elevation at $M_\star<10^{10.5}~M_\odot$ and lowering at $M_\star>10^{11.5}~M_\odot$ (not explicitly plotted here). Therefore, our qualitative results in Section~\ref{sec: channelgrow} remain unchanged in both the high-seed and zevo-seed cases.\par
\begin{figure}
\includegraphics[width=\hsize]{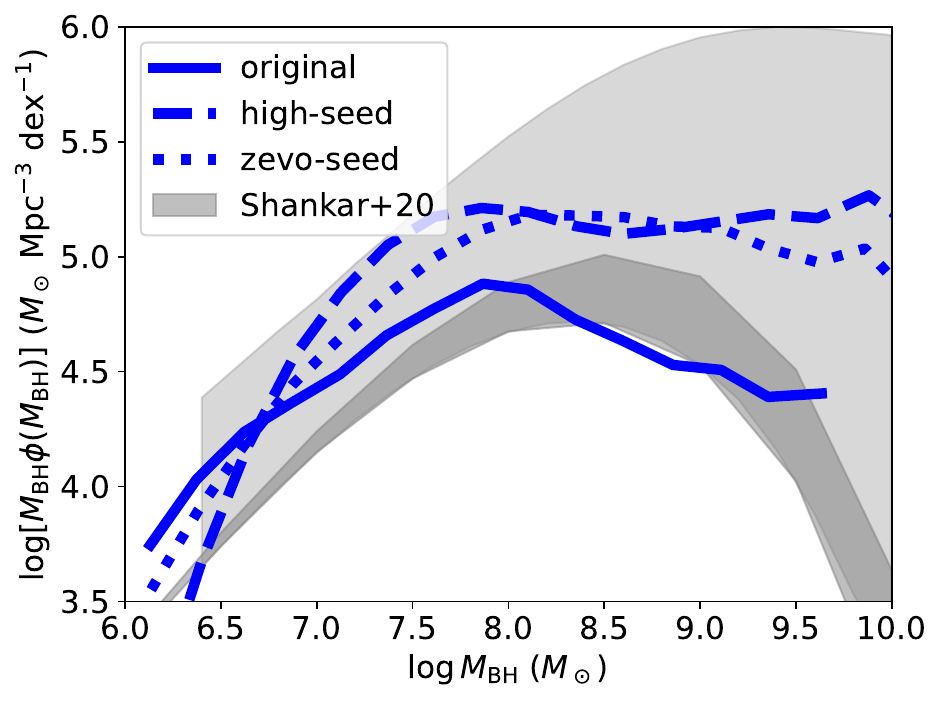}
\caption{The local BHMFs under different seeding cases, as labeled in the legend. The shaded region is from \citet{Shankar20b} for comparison, as explained in Figure~\ref{fig: bhmf_z0}.}
\label{fig: bhmf_z0_diffseed}
\end{figure}

However, the BHMF is more sensitive to the elevation of $M_\mathrm{BH}$. We show the local BHMFs in Figure~\ref{fig: bhmf_z0_diffseed}. Although not plotted, the new BHMFs would drop quickly at $M_\mathrm{BH}>10^{10}~M_\odot$. The deficiency of SMBHs with $M_\mathrm{BH}<10^7~M_\odot$ in the new BHMFs is caused by our $M_\star$ cut at $10^{9.5}~M_\odot$. Our new BHMFs have more SMBHs at the massive end than our original one and are also flatter, primarily because $M_\mathrm{BH}$ becomes systematically higher. Therefore, the new BHMFs become higher than the debiased BHMF in \citet{Shankar20b}. This reflects the additional uncertainty from the seeding, where the BHMF may be noticeably elevated if the seed mass becomes comparable to the accreted mass.

%%%%%%%%%%
% Summary
%%%%%%%%%%
\section{Summary}
\label{sec: summary}
In this work, we use a hybrid approach to combine the observationally measured SMBH accretion in \citet{Zou24} and the merger trees in the TNG simulations to probe the evolution of the SMBH population from $z=4$ to $z=0$. Taking advantage of both observations and simulations, we aim to construct the most realistic picture of cosmic SMBH evolution. Our main results are summarized as follows:
\begin{enumerate}
\item We reproduced the local $M_\mathrm{BH}-M_\star$ scaling relation and showed that half of the observed scatter of this scaling relation can be accounted for by SMBH growth along diverse $M_\star$ evolutionary tracks. Our relation is lower than the original TNG-simulated relation and also dynamically measured ones but is more consistent with those based on AGNs or after corrections for the selection bias of dynamically measured SMBHs, supporting the claim that the observed dynamically measured scaling relation may be biased. We predicted that the scaling relation does not have a strong redshift evolution, at least up to $z=3$, as primarily regulated by the correlation between $\dot{M}_a$ and $(M_\star, z)$. See Section~\ref{sec: scaling}.
\item We recovered the local BHMF and obtained a local SMBH mass density $\rho_\mathrm{BH}=1.5\times10^5~M_\odot~\mathrm{Mpc}^{-3}$. Our BHMF is lower than the original TNG-simulated ones and is more consistent with the one based on the debiased scaling relation. We predict that the BHMF steadily increases at all $M_\mathrm{BH}$ from $z=4$ to $z=1$ and largely remains frozen from $z=1$ to $z=0$. See Section~\ref{sec: bhmf}.
\item The overall SMBH growth is generally dominated by the accretion channel at $z\gtrsim1$. The relative contribution from mergers becomes increasingly important with rising $M_\star$ and decreasing redshift. $\dot{M}_m$ becomes comparable with $\dot{M}_a$ when $M_\mathrm{BH}\gtrsim10^8~M_\odot$ or $M_\star\gtrsim10^{11}~M_\odot$ at $z\lesssim1$. The exact transitioning $M_\mathrm{BH}$ or $M_\star$ from accretion-dominated growth to merger-dominated growth depends upon $p_\mathrm{merge}$ and contributions from MBHs in dwarf galaxies to mergers. See Section~\ref{sec: channelgrow}.
\item Half of $M_\mathrm{BH}$ is accumulated by $z\approx0.8-2$. Besides, SMBHs in massive galaxies with $M_\star\gtrsim10^{10}~M_\odot$ quickly lose their memories about their initial seeds at high redshift, while SMBHs in low-mass galaxies with $M_\star\lesssim10^{10}~M_\odot$ still retain information about their seed mass. See Section~\ref{sec: massassembly}.
\item Around 25\% of the SMBH mass budget is locked in long-lived wandering SMBHs, and the wandering mass fraction and wandering SMBH counts strongly increase with rising $M_\star$. See Section~\ref{sec: wander}.
\item We have examined the sensitivity of our results to the $p_\mathrm{merge}$ choice. The overall central-SMBH demography is robust against $p_\mathrm{merge}$ due to the limited contributions of mergers to the overall SMBH growth, but $p_\mathrm{merge}$ regulates quantities directly relevant to mergers, such as $\dot{M}_m$ and wandering SMBHs. See Section~\ref{sec: differentpmerge}.
\item We also tested different seeding choices. If the dynamically measured relation is adopted, or there is a strong positive redshift evolution of the scaling-relation normalization, our recovered $M_\mathrm{BH}-M_\star$ relation would be systematically elevated by $\approx0.3$~dex at $z\leq2$, and the massive end of the BHMF would have more noticeable difference. However, such a redshift evolution is not favored by observations (e.g., see Figure~10 in \citealt{Li23}), and the elevation requires SMBHs to accumulate around half of their total masses by $z=4$, which may also be questionable. See Section~\ref{sec: differentseed}.
\end{enumerate}
Overall, our results justify the power of the hybrid approach of combining observations and simulations. Further improvements will require future efforts in both directions. We are currently confined within $M_\star>10^{9.5}~M_\odot$ and $z<4$ by observations, and more sensitive observations, especially future deep \mbox{X-ray} surveys performed by the Athena, AXIS, and Lynx missions, are needed to push the $\dot{M}_a$ measurements down to the dwarf-galaxy regime and/or into the earlier universe. Another direction is to improve the simulations. The primary limitation, as illustrated in Figure~\ref{fig: bhar_tng} and discussed in more detail in \citet{Habouzit21, Habouzit22}, is that the simulated SMBH population is inconsistent with observations, and thus the SMBH modeling in current cosmological simulations should be improved. For example, at $z\gtrsim2$, the simulated $\dot{M}_a$ is systematically higher than for observations; while at $z\lesssim2$, the simulated $\dot{M}_a$ is too high at $M_\star\lesssim10^{11}~M_\odot$ but too low at $M_\star\gtrsim10^{11}~M_\odot$. Similarly, \citet{Weinberger18} and \citet{Habouzit22} showed that the TNG-simulated AGN luminosity functions are too high at high redshifts. These indicate that, for TNG, its Bondi accretion model and also its suppression of the AGN emission in massive galaxies due to the transitioning of SMBHs into the kinetic mode are overly efficient. Future simulations could use the observed AGN population properties to help calibrate their subgrid accretion physics. Besides, future work should also try to constrain the SMBH dynamics after galaxy mergers, such as the merger probability and the time lag between SMBH mergers and galaxy mergers, which can help better characterize the merger-driven growth channel. It is particularly beneficial to construct relationships between these SMBH dynamical parameters (including their distributions) and global galaxy properties.

\acknowledgments
We thank the anonymous referee for constructive suggestions and comments. We thank Lumen Boco and Andrea Lapi for providing the BHMF data. We thank Chris Byrohl, Kelly Holley-Bockelmann, Joel Leja, Dylan Nelson, Steinn Sigurdsson, and Guang Yang for helpful discussions. F.Z., W.N.B., and Z.Y. acknowledge financial support from NSF grant AST-2106990, Chandra X-ray Center grant AR4-25008X, the Penn State Eberly Endowment, and Penn State ACIS Instrument Team Contract SV4-74018 (issued by the Chandra X-ray Center, which is operated by the Smithsonian Astrophysical Observatory for and on behalf of NASA under contract NAS8-03060). B.L. acknowledges financial support from the National Natural Science Foundation of China grant 11991053. Y.X. acknowledges support from NSFC grants (12025303 and 12393814). The Chandra ACIS Team Guaranteed Time Observations (GTO) utilized were selected by the ACIS Instrument Principal Investigator, Gordon P. Garmire, currently of the Huntingdon Institute for X-ray Astronomy, LLC, which is under contract to the Smithsonian Astrophysical Observatory via Contract SV2-82024.

\appendix
\section{Effects of Statistical Uncertainties of the Accretion Rate}
\label{append: adderr}
\begin{figure}
\includegraphics[width=\hsize]{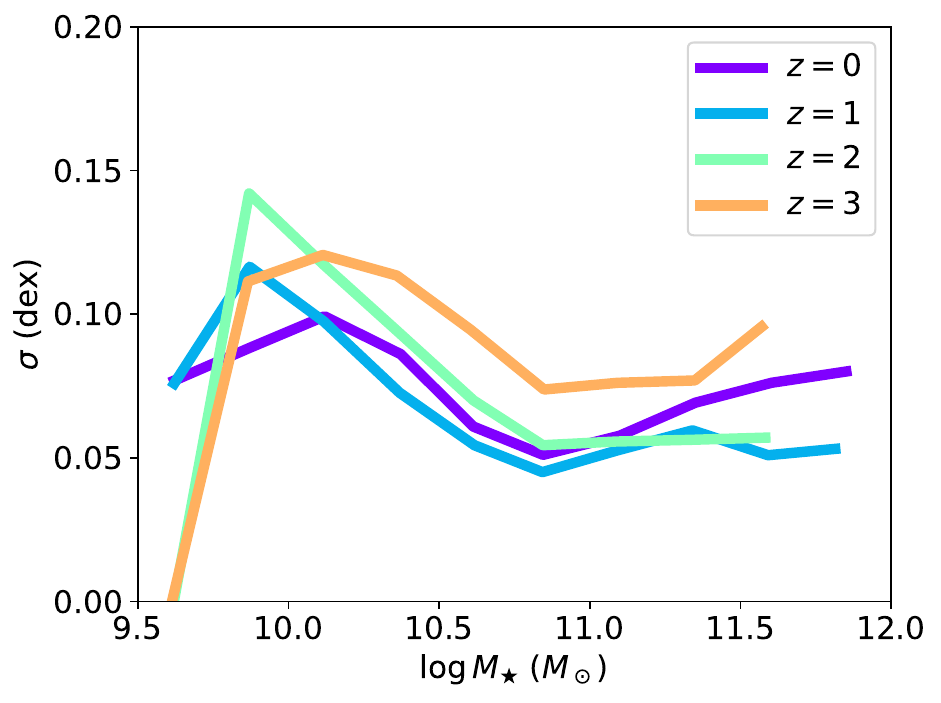}
\caption{The expected additional uncertainties of $M_\mathrm{BH}$ in dex from the statistical uncertainties of $\dot{M}_a$, color-coded by the redshift.}
\label{fig: adderr}
\end{figure}

In the main text, we do not add the statistical uncertainties of the $\dot{M}_a$ measurements in \citet{Zou24} because they do not reflect any physical scatters of the SMBH population. In principle, the statistical uncertainties would approach zero with an infinite amount of data. In this Appendix, we assess the level of additional uncertainties that can be induced into $M_\mathrm{BH}$ from the $\dot{M}_a$ measurements. We randomly draw a $\dot{M}_a(M_\star, z)$ function from the sampling results in \citet{Zou24} for each source and rerun our analyses in Section~\ref{sec: methodology}. We derive the additional uncertainties by comparing the new $M_\mathrm{BH}$ values with the original ones and subtracting the variance from the seeding scatter, as discussed when presenting Figure~\ref{fig: scaling} in Section~\ref{sec: scaling}. We show the $1\sigma$ uncertainty versus $M_\star$ in Figure~\ref{fig: adderr}. The figure shows that the additional uncertainty from the statistical uncertainties of $\dot{M}_a$ is only $\approx0.05-0.1$~dex across the whole parameter space, which should be smaller than possible systematic uncertainties. Additionally, this effect almost does not affect the median values of our results. Therefore, the statistical uncertainties of $\dot{M}_a$ are largely negligible for our $M_\mathrm{BH}$ measurements. Nevertheless, it should be noted that the elevated $\dot{M}_a$ uncertainties at $z\approx0$ become more important when comparing the relative contributions of $\dot{M}_a$ and $\dot{M}_m$ in Section~\ref{sec: channelgrow}.

\section{Comparisons between the TNG100 and TNG300 Results}
\label{append: comptng}
\begin{figure}
\includegraphics[width=\hsize]{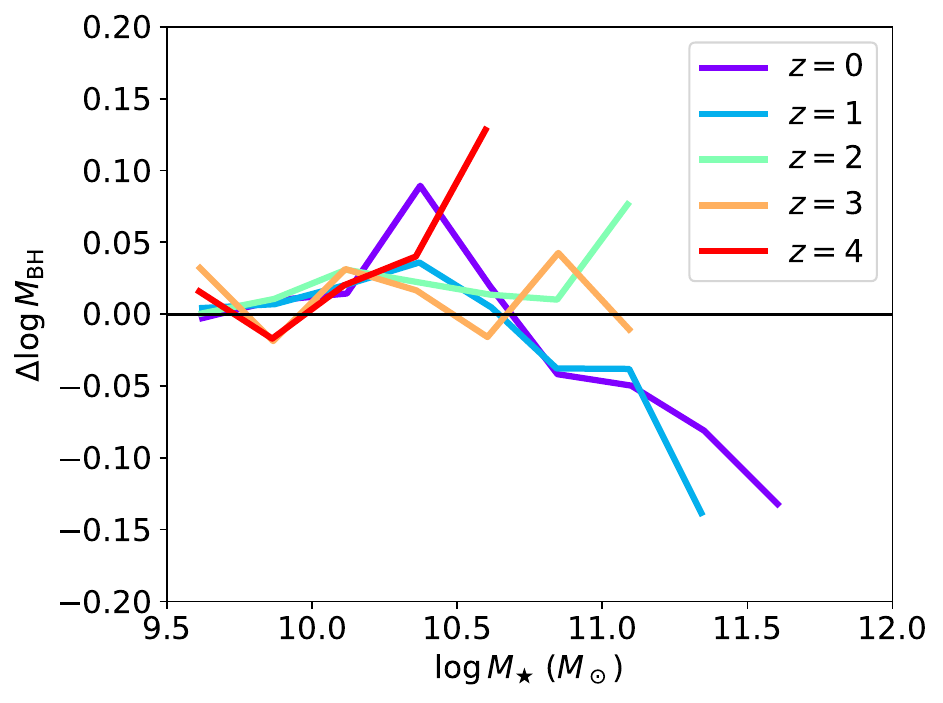}
\caption{The difference between the median TNG100 and TNG300 $\log M_\mathrm{BH}$ versus $M_\star$ at different redshifts. The difference is generally small, indicating that TNG100 and TNG300 $M_\mathrm{BH}-M_\star$ scaling relations are similar.}
\label{fig: detmbh_difftng}
\end{figure}

We focused on TNG300 in the main text, and this appendix compares the results based on TNG100 and TNG300. We show the difference between the TNG100 and TNG300 scaling relations in Figure~\ref{fig: detmbh_difftng}. The figure indicates that the two scaling relations are similar at all redshifts, with differences generally less than 0.05~dex. Similarly, we have compared all of our other analyses, and TNG100 results are also included in all of our tables for completeness. We do not explicitly plot all of them for conciseness, but the TNG100 and TNG300 results are very similar. The only more noteworthy difference is that the TNG300 SMF has a slightly heavier tail at the massive end than for TNG100, causing the mean TNG100 $M_\mathrm{BH}$ to shift downward by $0.1-0.2$~dex compared to TNG300. This effect can be seen more clearly in Table~\ref{tbl: massquantiles} but becomes insignificant when comparing parameters conditioned on a fixed $M_\star$.

\bibliography{citations}
\end{document}